\documentclass[twocolumn]{aastex63}
\pdfoutput=1
\usepackage{graphicx}
\usepackage{enumerate}
\usepackage{amssymb, amsmath, amsfonts}
\usepackage{gensymb}
\usepackage{mathtools}
\usepackage{natbib}
\usepackage{color}
\usepackage{hyperref}
\usepackage{ulem}
\usepackage{soul}
\usepackage{url}
\usepackage{float}
\usepackage{xspace}
\usepackage{comment}
\usepackage{longtable}
\usepackage{array}
\usepackage{lineno}
\usepackage{cellspace}
\newcommand{\Hipparcos}{{\sl Hipparcos }}
\newcommand{\Gaia}{{\sl Gaia }}

\newcommand{\Lsun}{\mbox{$L_{\sun}$}}

\newcommand{\Mjup}{\mbox{$M_{\rm Jup}$}}

\newcommand{\Lbol}{\mbox{$L_{\rm bol}$}}

\newcommand{\Teff}{\mbox{$T_{\rm eff}$}}
\newcommand{\logg}{\mbox{$\log(g)$}}

\newcommand{\masyr}{\hbox{mas\,yr$^{-1}$}}

\newcommand{\epsindi}{$\varepsilon$~Indi\xspace}
\newcommand{\epsindib}{$\varepsilon$~Indi~B\xspace}
\newcommand{\epsindiba}{$\varepsilon$~Indi~Ba\xspace}
\newcommand{\epsindibb}{$\varepsilon$~Indi~Bb\xspace}
\newcommand{\Ba}{Ba\xspace}
\newcommand{\Bb}{Bb\xspace}
\newcommand{\MBa}{\hbox{$66.92\pm 0.36$}}
\newcommand{\MBb}{\hbox{$53.25\pm 0.29$}}
\newcommand{\MB}{\hbox{$120.17\pm 0.62$}}

\newcommand{\RomanNumeralCaps}[1]
    {\MakeUppercase{\romannumeral #1}}

\begin{document}

\title{Precise Dynamical Masses of $\varepsilon$ Indi Ba and Bb: Evidence of Slowed Cooling at the L/T Transition}

 \author[0000-0001-8892-4045]{Minghan Chen}
\affiliation{Department of Physics, University of California, Santa Barbara, Santa Barbara, CA 93106, USA}
\affiliation{Joint first author}

\author[0000-0002-6845-9702]{Yiting Li}
\affiliation{Department of Physics, University of California, Santa Barbara, Santa Barbara, CA 93106, USA}
\affiliation{Joint first author}

\author[0000-0003-2630-8073]{Timothy D.~Brandt}
\affiliation{Department of Physics, University of California, Santa Barbara, Santa Barbara, CA 93106, USA}

\author[0000-0001-9823-1445]{Trent J.~Dupuy}
\affiliation{Institute for Astronomy, University of Edinburgh, Royal Observatory, Blackford Hill, Edinburgh, EH9 3HJ, UK}

\author[0000-0000-0000-0000]{C\'atia V.~Cardoso}
\affiliation{European Space Agency, Keplerlaan 1, PO Box 299 NL-2200 AG Noordwijk, The Netherlands}

\author[0000-0002-1452-5268]{Mark J.~McCaughrean}
\affiliation{European Space Agency, Keplerlaan 1, PO Box 299 NL-2200 AG Noordwijk, The Netherlands}

\begin{abstract}
We report individual dynamical masses of \MBa$\,\Mjup$ and \MBb$\,\Mjup$ for the binary brown dwarfs \epsindiba and \Bb, measured from long term ($\approx$10\,yr) relative orbit monitoring and absolute astrometry monitoring data on the VLT. Relative astrometry with NACO fully constrains the Keplerian orbit of the binary pair, while absolute astrometry with FORS2 measures the system's parallax and mass ratio.  We find a parallax consistent with the Hipparcos and Gaia values for \epsindi~A, and a mass ratio for \epsindiba to \Bb precise to better than 0.2\%. \epsindiba and \Bb have spectral types T1-1.5 and T6, respectively. With an age of $3.5_{-1.0}^{+0.8}$\,Gyr from \epsindi~A's activity, these brown dwarfs provide some of the most precise benchmarks for substellar cooling models.  Assuming coevality, the very different luminosities of the two brown dwarfs and our moderate mass ratio imply a steep mass-luminosity relationship $L \propto M^{5.37\pm0.08}$ that can be explained by a slowed cooling rate in the L/T transition, as previously observed for other L/T binaries. Finally, we present a periodogram analysis of the near-infrared photometric data, but find no definitive evidence of periodic signals with a coherent phase. 
\end{abstract}


\section{Introduction} \label{sec:intro}

Brown dwarfs (BDs) are sub-stellar objects below the hydrogen burning limit ($\lesssim$80\;$\Mjup$) but massive enough to fuse deuterium ($\gtrsim$13\;$\Mjup$) \citep{Spiegel_2011,Dieterich_2014}. After their formation, BDs cool radiatively and follow mass-luminosity-age relationships. The degeneracy in these parameters, especially in mass and age, plus assumptions about the initial conditions of BD formation, have long been a major difficulty in calibrating the evolutionary and atmospheric models for substellar objects \citep{Burrows_1989,Baraffe_2003,Joergens_2006,Gomes_2012,Helling_2014,Caballero_2018}. BDs for which we can measure these parameters independently can benchmark the evolutionary models. As a result, BDs in multiple systems are especially important.  Their ages can be determined from the characteristics of their host stars or associated groups, assuming coevality \citep[e.g.][]{Seifahrt_2010,Leggett_2017}, and some of these BDs can have their masses measured dynamically \citep[e.g.][]{Crepp+Johnson+Fischer+etal_2012,Crepp+Gonzales+Bechter+etal_2016,Dupuy+Liu_2017,Dieterich_2018,Brandt_2019,Brandt_2020}. 

\epsindib, discovered by \cite{Scholz_2003_EpsIndiB_discovery}, is a distant companion to the high proper motion ($\sim$4.7 arcsec/yr) star \epsindi. It was later resolved to be a binary brown dwarf system by \cite{McCaughrean_2004}, who estimated the two components of the binary, \epsindiba and \Bb, to be T dwarfs with spectral types T1 and T6, respectively. It was the first binary T dwarf to be discovered and remains one of the closest binary brown dwarf systems to our solar system; \Gaia EDR3 measured a distance of $3.638 \pm 0.001$\,pc to \epsindi~A \citep{Lindegren+Klioner+Hernandez+etal_2020}. Their proximity makes \epsindiba and \Bb bright enough and their projected separation wide enough to obtain high quality, spatially resolved images and spectra. And their relatively short orbital period of $\approx$10\,yr allows the entire orbit to be traced in a long-term monitoring campaign. Being near the boundary of the L-T transition, \epsindiba is especially valuable for understanding the atmospheres of these ultra-cool brown dwarfs \citep{Apai_2010,Goldman_2008,Rajan_2015}.

\cite{King_2010} carried out a detailed photometric and spectroscopic study of the system, and derived luminosities of $\log_{10} L/L_{\odot} = -4.699 \pm 0.017$ and $-5.232 \pm 0.020$ for \Ba and \Bb, respectively. They found that neither a cloud-free nor a dusty atmospheric model can sufficiently explain the brown dwarf spectra, and that a model allowing partially settled clouds produced the best match. The relative orbit monitoring was still ongoing at the time, so a preliminary dynamical system mass of $121 \pm 1\,\Mjup$ measured by \citet{Cardoso_2012} was assumed by the authors to derive mass ranges of 60-73 $\Mjup$ and 47-60 $\Mjup$ for \Ba and \Bb based on their photometric and spectroscopic observations.

\cite{Cardoso_2012} and \cite{Dieterich_2018} both used a combination of absolute and relative astrometry to obtain individual dynamical masses of
\epsindiba and \Bb.
\cite{Cardoso_2012} used NACO \citep{NACO_ins_paper_1,NACO_ins_paper_2} and FORS2 \citep{Appenzeller+Fricke+Furtig+etal_1998,FORS2_ADC_1997} imaging to measure $77.8\pm0.3$\,$M_{\rm Jup}$ and $61.9\pm0.3$\,$M_{\rm Jup}$ for \epsindiba and \Bb, respectively, with a parallax of $263.3\pm 0.3$\,mas. This parallax disagreed strongly with the \Hipparcos parallax { of \epsindi~A} \citep{ESA_1997,vanLeeuwen_2007}. Fixing parallax to the \Hipparcos 2007 value of $276.1 \pm 0.3$\,mas, \cite{Cardoso_2012} instead obtained masses of $68.0\pm0.9$\,$M_{\rm Jup}$ and $53.1\pm0.3$\,$M_{\rm Jup}$.  \cite{Dieterich_2018} used a different data set to measure individual masses of $75.0\pm0.8$\,$M_{\rm Jup}$ and $70.1\pm0.7$\,$M_{\rm Jup}$ with a parallax of $276.9\pm0.8$\,mas, consistent with the \Hipparcos distance.  The three dynamical mass measurements---two from \cite{Cardoso_2012} and one from \cite{Dieterich_2018}---disagree strongly with one another. The highest masses of $\gtrsim$75\,$M_{\rm Jup}$ are in tension with the predictions of substellar cooling models even at very old ages \citep{Dieterich_2014}.

In this paper, we use relative orbit and absolute astrometry monitoring of \epsindib from 2005 to 2016 acquired with the VLT to measure the individual dynamical masses of \epsindiba and \Bb. Much of this data set overlaps with that used by \cite{Cardoso_2012}, but we have the advantage of a few more epochs of data, \Gaia  astrometric references \citep{Lindegren+Klioner+Hernandez+etal_2020} and a better understanding of the direct imaging system thanks to years of work on the Galactic center \citep{Gillessen+Eisenhauer+Trippe+etal_2009,Plewa+Gillessen+Eisenhauer+etal_2015,Gillessen+Plewa+Eisenhauer+etal_2017}. We structure the paper as follows. We review the stellar properties and its age in Section \ref{sec:stellarprop}.  Section \ref{sec:data} presents the VLT data that we use, and Section \ref{sec:positions} describes our method for measuring calibrating the data and measuring the position of the two BDs.  Section \ref{sec:photvar} presents our search for periodic photometric variations, while in Section \ref{sec: orbit fit} we fit for the orbit and mass of the pair.  In Section \ref{sec:BDtests} we discuss the implications of our results for models of substellar evolution.  We conclude with Section \ref{sec:conclusions}.

\section{Stellar Properties} \label{sec:stellarprop}

The \epsindib system is bound to \epsindi A (=HIP 108870, HD 209100, HR 8387), a bright K4V or K5V star \citep{Adams_1935,Evans_1957,Gray_2006}. \epsindi A has a $2.7^{+2.2}_{-0.4}\,\Mjup$ planet on a low eccentricity and wide orbit \citep{Endl+Kurster+Els+etal_2002,Zechmeister+Kurster+Endl+etal_2013,Feng_2019}. The star appears to be slightly metal poor.  Apart from a measurement of ${\rm [Fe/H]} = -0.6$\,dex \citep{Soto+Jenkins_2018}, literature spectroscopic measurements range from ${\rm [Fe/H]} = -0.23$\,dex \citep{Abia+Rebolo+Beckman+etal_1988} to $+0.04$\,dex \citep{Kollatschny_1980}, with a median of $-0.17$\,dex \citep{Soubiran_2016}. 

Several studies have constrained the age of the \epsindi system via various methods such as evolutionary models, Ca\,{\sc ii}~HK age dating techniques, and kinematics. Using a dynamical system mass of $121 \pm 1 \Mjup$ and evolutionary models, \citet{Cardoso_2012} predicted a system age of $3.7$-$4.3$\,Gyr. This age is older than the age of $0.8$-$2.0$\,Gyr derived from stellar rotation {of \epsindi A} and the age of $1$-$2.7$\,Gyr from the Ca \RomanNumeralCaps{2} activity {of \epsindi A}, reported in \cite{Lachaume_1999} { assuming a stellar rotation of $\sim$20 days}, but is younger than the kinematic estimate of $>$7.4\,Gyr quoted in the same study. { \citet{Feng_2019} inferred a longer rotation period of $\sim$35 days derived from a relatively large data set of high precision RVs and multiple activity indicators for \epsindi A, and found an age of $\sim$4 Gyr}. To date, the age of the star still remains a major source of uncertainty in the evolutionary and atmospheric modeling of the system. 

\begin{figure}
    \centering
    \includegraphics[width=0.45\textwidth]{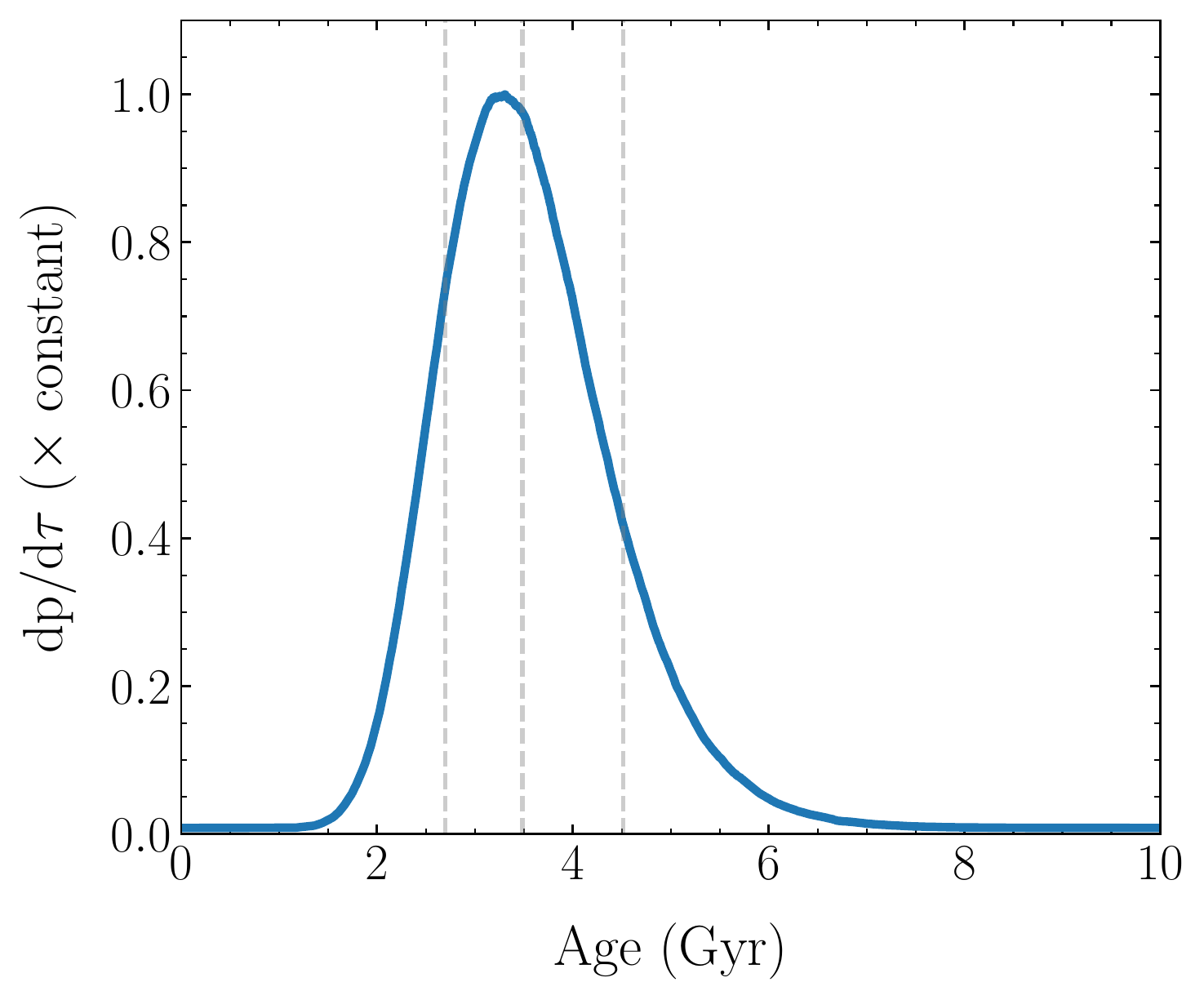}
    \caption{Age posterior of \epsindi A based on the Bayesian activity-age method of \citet{Brandt_2014}. Our analysis does not use a directly measured rotation period for \epsindi. The median and 1$\sigma$ uncertainties are shown by the grey dotted lines; they correspond to $3.48_{-1.03}^{+0.78}$ Gyr.}
    \label{fig::age}
\end{figure}

We perform our own analysis on the age of \epsindi using a Bayesian activity-based age dating tool devised by \citet{Brandt_2014} and applied in \cite{Li+Brandt+Brandt+etal_2021}. To do this, we adopt a Ca\,{\sc ii} chromospheric index of $\log R'_{\rm HK} = -4.72$ from \citet{Pace_2013}, an X-ray activity index of $R_{X} = -5.62$ from the ROSAT all-sky survey bright source catalog \citep{Voges_1999}, and Tycho $B_T V_T$ photometry ($B_T = 6.048\pm0.014$\,mag, $V_T = 4.826\pm0.009$\,mag) from the Tycho-2 catalog \citep{Hog+Fabricius+Makarov+etal_2000}. The star lacks a published photometric rotation period. Figure \ref{fig::age} shows our resulting posterior probability distribution, with an age of $3.48_{-1.03}^{+0.78}$ Gyr. This age is somewhat older than the young ages most literature measurements suggest, but is similar to the system age of $3.7$-$4.3$\,Gyr used by \citet{Cardoso_2012} for their analysis { based on the  preliminary system mass for \epsindi{} Ba+Bb compared to evolutionary models} { and to the $\sim$4\,Gyr age more recently inferred by \cite{Feng_2019}}. We use our Bayesian age posterior when analyzing the consistency with our dynamical masses with brown dwarf models (Section \ref{sec:BDtests}).

\section{Data} \label{sec:data}

\subsection{Relative Astrometry}

We measure the relative positions of \epsindiba and \Bb using nine years of monitoring by the Nasmyth Adaptive Optics System (NAOS) + Near-Infrared Imager and Spectrograph (CONICA), NACO for short \citep{NACO_ins_paper_1, NACO_ins_paper_2}. We use images taken by the S13 Camera on NACO in the $J$, $H$ and $K_s$ passbands. 
Our images come from Program IDs 072.C-0689(F), 073.C-0582(A), 074.C-0088(A), 075.C-0376(A), 076.C-0472(A), 077.C-0798(A), 078.C-0308(A), 079.C-0461(A), 380.C-0449(A), 381.C-0417(A), 382.C-0483(A), 383.C-0895(A), 384.C-0657(A), 385.C-0994(A), 386.C-0376(A), 087.C-0532(A), 088.C-0525(A), 089.C-0807(A), and 091.C-0899(A), all PI McCaughrean, and 381.C-0860(A), PI Kasper.

The S13 camera on NACO has a field of view (FOV) of $14'' \times 14''$ and a plate scale of $\approx$13.2$\,{\rm mas\,pix^{-1}}$. Most observing sequences consisted of $\approx$5 dithered images in each filter. The binary system HD~208371/2 was usually observed on the same nights and in the same mode to serve as an astrometric calibrator.  We use a total of 939 images of \epsindiba and \Bb, taken over 56 nights of observations from 2004 to 2013 for which we have contemporaneous imaging of HD~208371/2.

We perform basic calibrations on all of these images.  For each night, we use contemporaneous dark images to identify bad pixels and to remove static backgrounds.  We construct and use a single, master flat field for all images.  We mask pixels for which the flatfield correction deviates by more than 20\% from its median or for which the standard deviation of the dark frames is more than five times its median standard deviation.  We then subtract the median dark image and divide by the flatfield image.

The data quality varies depending on the observing conditions and the performance of the adaptive optics (AO) system. Therefore, we apply a selection criterion to exclude poor quality data. We first extract the sources in the images using the DAOPHOT program as implemented in the {\tt photutils} python package \citep{Stetson_1987, photutils110}. We obtain estimates of the following parameters for \epsindiba and \Bb: centroid, sharpness (a DAOPHOT parameter that characterizes the width of the source), roundness (a DAOPHOT parameter that characterizes the symmetry of the source), and signal to noise ratio (SNR). We discard images where one or both of the two targets fall outside the field of view, and for the remaining images we apply the following cut-offs in DAOPHOT detection parameters to exclude highly extended, highly elongated and low signal to noise images: sharpness $\geqslant 0.3$, $- 0.5 \leqslant $ roundness $\leqslant 0.5$ and SNR $\geqslant 25$. We then visually inspect the remaining images to remove ones with bad pixels (cosmic rays or optical defects) landing on or near the target objects, and ones with AO correction artifacts that survived our DAOPHOT cut. Table \ref{table:relastrodata} summarizes the final data set selected for relative astrometry measurements.

\begin{deluxetable}{cccc}
\tablewidth{0pt}
\tablecaption{Relative astrometry data summary \label{table:relastrodata} }
\tablehead{
Date & 
Filter(s) &
\# Frames &
Total integration (s)
}
\startdata
     2004-09-24 &  $J$, $H$, $K_s$ &       15 & 150 \\
     2004-11-14 &  $J$, $H$, $K_s$ &       14 & 840 \\
     2004-11-15 &        $J$ &        5 & 270 \\
     2004-12-15 &     $J$, $H$ &       11 & 220 \\
     2005-06-04 &  $J$, $H$, $K_s$ &       13 & 780 \\
     2005-07-06 &        $K_s$ &        6 & 310 \\
     2005-08-06 &  $J$, $H$, $K_s$ &       13 & 780 \\
     2005-12-17 &     $J$, $K_s$ &        7 & 210 \\
     2005-12-30 &  $J$, $H$, $K_s$ &       14 & 840 \\
     2005-12-31 &  $J$, $H$, $K_s$ &       13 & 780 \\
     2006-07-19 &     $H$, $K_s$ &        8 & 80 \\
     2006-08-06 &     $J$, $H$ &       10 & 100 \\
     2006-09-22 &  $J$, $H$, $K_s$ &       15 & 150 \\
     2006-10-03 &     $J$, $H$ &        7 & 420 \\
     2006-10-20 &     $J$, $H$ &        5 & 300 \\
     2006-11-12 &        $J$ &        5 & 300 \\
     2007-06-18 &  $J$, $H$, $K_s$ &       12 & 720 \\
     2007-09-09 &     $J$, $H$ &       10 & 450 \\
     2007-09-29 &     $J$, $H$ &       15 & 900 \\
     2007-11-07 &     $J$, $H$ &       10 & 600 \\
     2008-06-05 &     $J$, $H$ &       10 & 600 \\
     2008-06-10 &     $J$, $H$ &        7 & 70 \\
     2008-06-21 &     $J$, $H$ &       10 & 100 \\
     2008-08-25 &     $J$, $H$ &        9 & 540 \\
     2008-12-01 &     $J$, $H$ &       12 & 720 \\
     2009-06-17 &  $J$, $H$, $K_s$ &       12 & 720 \\
     2010-08-01 &     $J$, $H$ &        7 & 105 \\
     2010-11-07 &     $J$, $H$ &       10 & 300 \\
     2011-07-18 &  $J$, $H$, $K_s$ &       13 & 390 \\
     2012-07-18 &     $J$, $H$ &        9 & 540 \\
     2012-09-14 &     $J$, $H$ &        9 & 540 \\
     2013-06-07 &     $J$, $H$ &       10 & 600
\enddata
\end{deluxetable}

\subsection{Absolute Astrometry} \label{subsec:AbsAstData}

The long term absolute position of \epsindib was monitored with the FOcal Reducer and low dispersion Spectrograph \citep[FORS,][]{Appenzeller+Fricke+Furtig+etal_1998} installed on ESO's UT1 telescope at the Very Large Telescope (VLT). The FORS system consists of twin imagers and spectrographs FORS1 and FORS2, collectively covering the visual and near UV wavelength. The absolute astrometry monitoring was done with the FORS2 imager coupled with two mosaic MIT CCDs; the camera has a pixel scale of 0$.\!\!''$126/pixel in its unbinned mode and a field of view (FOV) of $\approx$8$.\!'6\times8.\!'6$. 

The FORS2 monitoring of \epsindib covers a long temporal baseline beginning in 2005 and ending in 2016. Our images come from Program IDs 072.C-0689(D), 075.C-0376(B), 076.C-0472(B), 077.C-0798(B), 078.C-0308(B), 079.C-0461(B), 380.C-0449(B), 381.C-0417(B), 382.C-0483(B), 383.C-0895(B), 384.C-0657(B), 385.C-0994(B), 386.C-0376(B), 087.C-0532(B), 088.C-0525(B), 089.C-0807(B), and 091.C-0899(B), all PI McCaughrean.
The FORS-2 focal plane consists of two CCDs, chip1 and chip2. We only consider the data taken with the chip1 CCD. Over the 12 years of absolute position monitoring, 940 images were taken with chip1 over 88 epochs. For the majority of the epochs, 10 dithered images in $I_{\rm BESSEL}$ filter were obtained, with a 20 second exposure time for each image. We exclude 36 blank image frames over 4 epochs between 2009-8-21 and 2009-11-3, resulting in a final total of 904 image frames for our analysis. A summary of the FORS2 data is given in Table \ref{tab:absast_obs_log}.
These 904 science frames are bias-corrected and flat-fielded using the normalized master values generated from median combination of the flat and bias frames obtained in the same set of observing programs. 

\begin{deluxetable}{cccc}
\tablewidth{0pt}
\tablecaption{Absolute astrometry data from FORS2\tablenotemark{a} \label{tab:absast_obs_log}}
\tablehead{Date & \# Frames & Band &
Total integration (s)}
\tablewidth{0pt}
\startdata
2005-05-06  & 10 & $I_{\rm Bess}$ & 200 \\
2005-05-12  & 10 & $I_{\rm Bess}$ &200 \\
2005-06-08  & 10 & $I_{\rm Bess}$ &200 \\
2005-07-06  & 10 & $I_{\rm Bess}$ &200
\enddata
\tablenotetext{a}{The full observing log is available as an online table; only the first four rows are shown here for reference.}
\end{deluxetable}

\section{Relative and Absolute Positions} \label{sec:positions}

\subsection{Point Spread Function (PSF) Fitting} \label{subsec:joint psf fit}

To measure the relative separations of the two brown dwarfs in the NACO data, we need to fit their PSFs. \cite{Cardoso_2012} has demonstrated that Moffat is the best analytical profile for the NACO data compared to Lorentzian and Gaussian. During the epochs when the projected separations of the two brown dwarfs are small, the two PSFs are only separated by one or two full widths at half maximum (FWHM). As a result, the flux near the center of one source has non-negligible contributions from the wings of the other source. This could introduce significant biases in the measured positions if fitting a PSF profile to each source separately. Therefore, we implement a joint fit of the two PSFs using a sum of two elliptical Moffat profiles:
\begin{equation}
    \label{eqn:joint moffat model}
{\rm Counts}(x, y) = f_1\psi_{1}(x, y) + f_2\psi_{2}(x, y)
\end{equation}
with
\begin{multline}
\label{eqn:elliptical moffat}
    \psi_{i}(x, y) = f_i (1 + c_1(x - x_i)^2 + 2c_2(x - x_i)(y - y_i) \\ + c_3(y - y_i)^2 ) ^{-\beta}
\end{multline}
where $\psi_i$ is a general elliptical 2D Moffat profile centered at \{$x_i, y_i$\} with peak intensity $f_i$.  Our model is the sum of two such profiles with different fluxes at different locations, sharing the same morphology, i.e., the same \{$c_1, c_2, c_3$\}. Instead of fitting for \{$c_1, c_2, c_3$\} directly, we fit for three equivalent parameters: \{${\rm fwhm}_x, {\rm fwhm}_y, \phi$\}, which are the FWHMs of the elliptical Moffat profile along the x and y axes, and the counter-clockwise rotation angle of the PSF, respectively. These physical parameters are related to \{$c_1, c_2, c_3, \beta$\} through the following equations:
\begin{align}
    c_1 &= \frac{\cos^2\phi}{\sigma_x^2} + \frac{\sin^2\phi}{\sigma_y^2}\\
    c_2 &= \frac{\sin 2\phi}{2\sigma_x^2} - \frac{\sin 2\phi}{2\sigma_y^2}\\
    c_3 &= \frac{\sin^2\phi}{\sigma_x^2} + \frac{\cos^2\phi}{\sigma_y^2}\\
    {\rm fwhm}_{x, y} &= 2\sigma_{x, y} \sqrt{(2^{1/\beta} - 1)}
\end{align} 

For each background subtracted image, we fit for the sum of two PSFs by minimizing $\chi^2$ over 10 parameters: \{$x_1, y_1, x_2, y_2, f_1, f_2, {\rm fwhm}_x, {\rm fwhm}_y, \phi, \beta$\}. In this case, $\chi^2$ is defined by:
\begin{align}
\label{eqn: moffat chisqr}
\chi^2 = \sum_{i}^{n_{pix}} \frac{(D_{i} - f_1\; \psi_{1, i} - f_2\; \psi_{2, i})^2}{\sigma_{i}^2}
\end{align}
We use scipy's non-linear optimization routines \citep{2020SciPy-NMeth} to minimize $\chi^2$ over the 8 non-linear parameters \{$x_1, y_1, x_2, y_2, {\rm fwhm}_x, {\rm fwhm}_y, \phi, \beta$\}, and for each trial set of non-linear parameters, we solve for the best fit linear parameters \{$f_1, f_2$\} analytically and marginalize over them.

\subsection{Calibrations for Relative Astrometry} \label{subsec: relast calibrations}

In order to measure precise relative astrometry, we must measure and correct various instrumental properties and atmospheric effects that can alter the apparent separation and position angle (PA) of \epsindiba and \Bb. In this section we describe our calibrations for the instrument plate scale and orientation, distortion correction and differential atmospheric refraction.

\subsubsection{Plate scale, Orientation, and Distortion Correction} \label{subsubsec: platecal}

We calibrate the plate scale and the north pointing of the NACO S13 camera using NACO's observations of a nearby wide separation binary, HD 208371/2, observed concurrently with the science data over the $\sim$10-year relative orbit monitoring period. We calibrate the separation and PA of the binary in NACO data against the high precision measurements from Gaia EDR3 for HD 208371/2:
\begin{align}
    \label{eq:EDR3 AB sep}
    \frac{\rm sep}{\rm arcsec} &= 8.90612 + 0.00011 \left({\rm Jyear} - 2016.0 \right) \\
    \label{eq:EDR3 AB PA}
    \frac{\rm PA}{\rm degree} &= 348.10345 - 0.00040 \left({\rm Jyear} - 2016.0 \right)
\end{align}
The uncertainties on these do depend on the epoch, but with proper motion uncertainties $\lesssim$40\,$\mu$as\,yr$^{-1}$, positional uncertainties are only $\approx$0.5\,mas even extrapolated ten years before Gaia.  This represents a fractional uncertainty in separation below 10$^{-4}$ and contributes negligibly to our error budget.

\begin{figure}
    \centering
    \includegraphics[width=\linewidth]{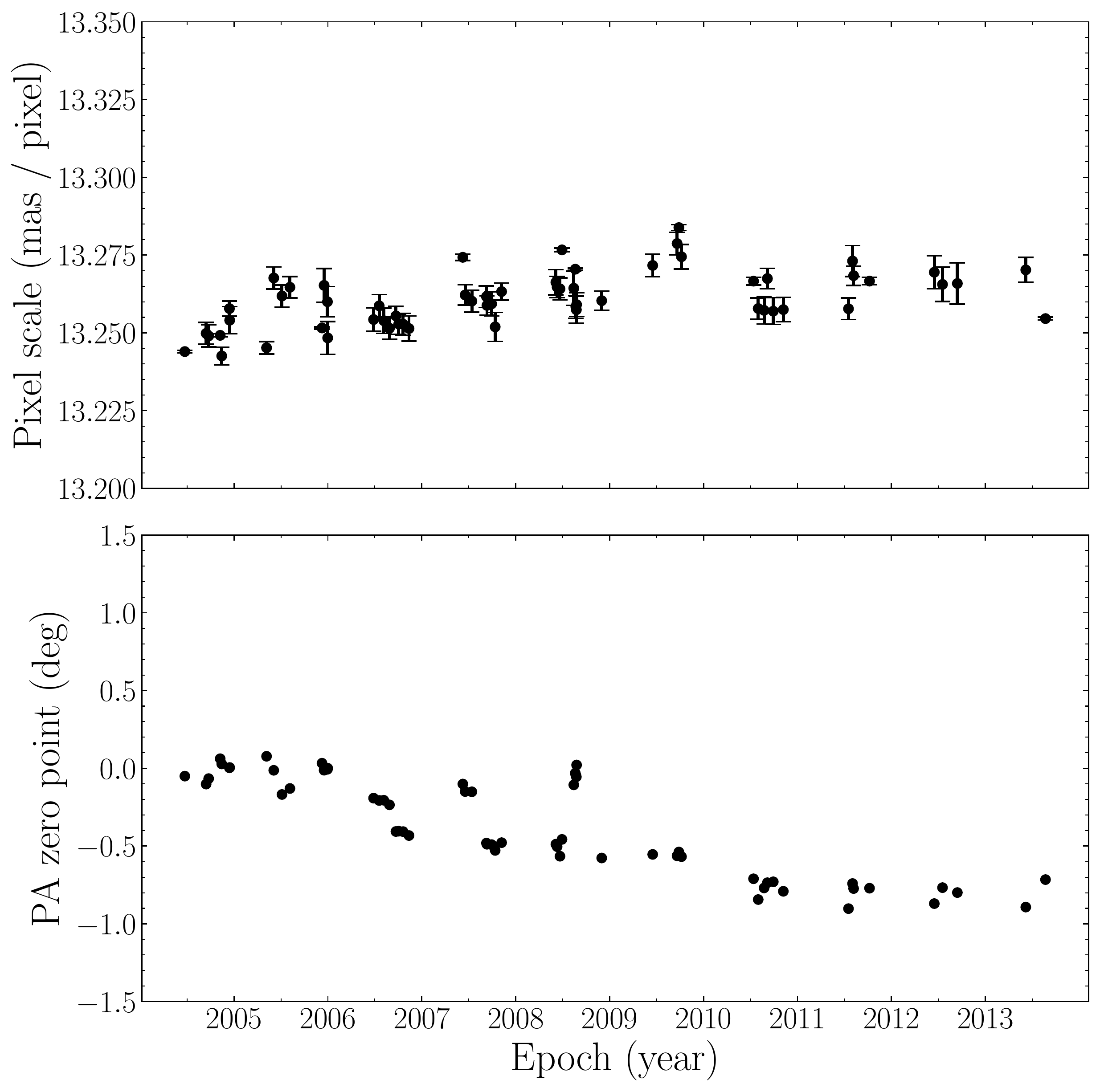}
    \caption{Pixel scale and PA zero point calibrations for the NACO S13 camera, derived using the binary HD~208371/2 as measured by Gaia EDR3.}
    \label{fig:naco calibrations}
\end{figure}

To measure the separation and PA of the calibration binary, we use the Moffat PSF fitting algorithm described in section \ref{subsec:joint psf fit}. Since the binary is widely separated, a joint PSF fit in this case is effectively equivalent to fitting a single 2D Moffat profile for each star separately {(albeit with the same structure parameters for each star's Moffat function)}. The calibration results are shown in Fig.\ref{fig:naco calibrations}. We measure an overall average plate scale of $13.260 \pm 0.001$, but we also note that the plate scale seems to increase slightly with time from 2004 to 2010. Both the plate scale and the increasing trend agree with other measurements in the literature, \citep{Chauvin_2010,Cardoso_2012}. In \cite{Cardoso_2012}, the same calibration binary was used to derive the plate scales but a different reference measurement for the binary was used. Adjusting their results to the more precise Gaia measurement of the binary brings their plate scale into agreement with ours. The PA zero point of the instrument varies from observation to observation, and has a long term trend as well. This is in agreement with the analysis in \cite{Plewa_2018}.

The distortion correction was shown to be of little significance for the NACO S13 camera \citep{Trippe_2008} due to the small field of view. For completeness, we still apply the distortion correction derived by \citet{Plewa_2018_distortion_correction}.

\subsubsection{Differential Atmospheric Refraction and Annual Aberration}
The dominant atmospheric effect that needs to be corrected for is differential atmospheric refraction \citep{Gubler_1998}. When a light ray travels from vacuum into Earth's atmosphere, it is refracted along the zenith direction and changes the observed zenith angle of the source, making the apparent zenith angle, $z$, deviate from the true zenith angle in the absence of an atmosphere, $z_0$:
\begin{equation}
    z = z_0 + R
\end{equation}
where R is the total refraction angle experienced by the light ray. The amount of this refraction depends on atmospheric conditions, the wavelength of the incoming light, and the zenith angle of the object. Therefore, for two objects at different positions in the sky and with different spectral types, the total refraction angles are different and can alter the apparent separation and PA of the objects. We can write this differential refraction, $\Delta R$, in terms of two components, one due to their difference in color, and one due to the difference of their true zenith angles \citep{Gubler_1998}:
\begin{equation}
    \Delta R = \Delta R_{\rm color} + \Delta R_{\Delta z_0}
\end{equation}
For \epsindiba and \Bb, the second term is {much smaller} as they are separated by only $< 1''$, {and produced negligible effects on the final results compared to the first term. We included both effects for completeness. The total differential refraction can be calculated with:}
\begin{equation}
    \label{eq: ADR}
    \Delta R = R_2(n_2, z_2) - R_1(n_1, z_1)
\end{equation}
where the $n_i$'s are the effective refractive indices of the Earth's atmosphere for the target sources. $n_i$ depends on the effective central wavelength ($\lambda_i$) of the target in the observed passband, and on observing conditions, most commonly pressure ($P$), temperature ($T$), humidity ($H$) and altitude ($z$). \citet{Cardoso_2012} calculated the effective central wavelengths for \epsindiba and \Bb in the $J$, $H$, and $K_s$ bands by integrating high resolution spectra of the two brown dwarfs. To calculate the refractive index, $n_i(\lambda_i, P, T, H, z)$, we use the models in \citet{Mathar_2007} covering a wavelength range of $1.3\;\mu$m to $24\; \mu$m. Then, the total refraction can be approximately expressed as \citep{smart_1977}:
\begin{equation}
    R(n, z) \approx (n - 1) \tan(z)
\end{equation}

\begin{figure}
    \centering
    \includegraphics[width=\linewidth]{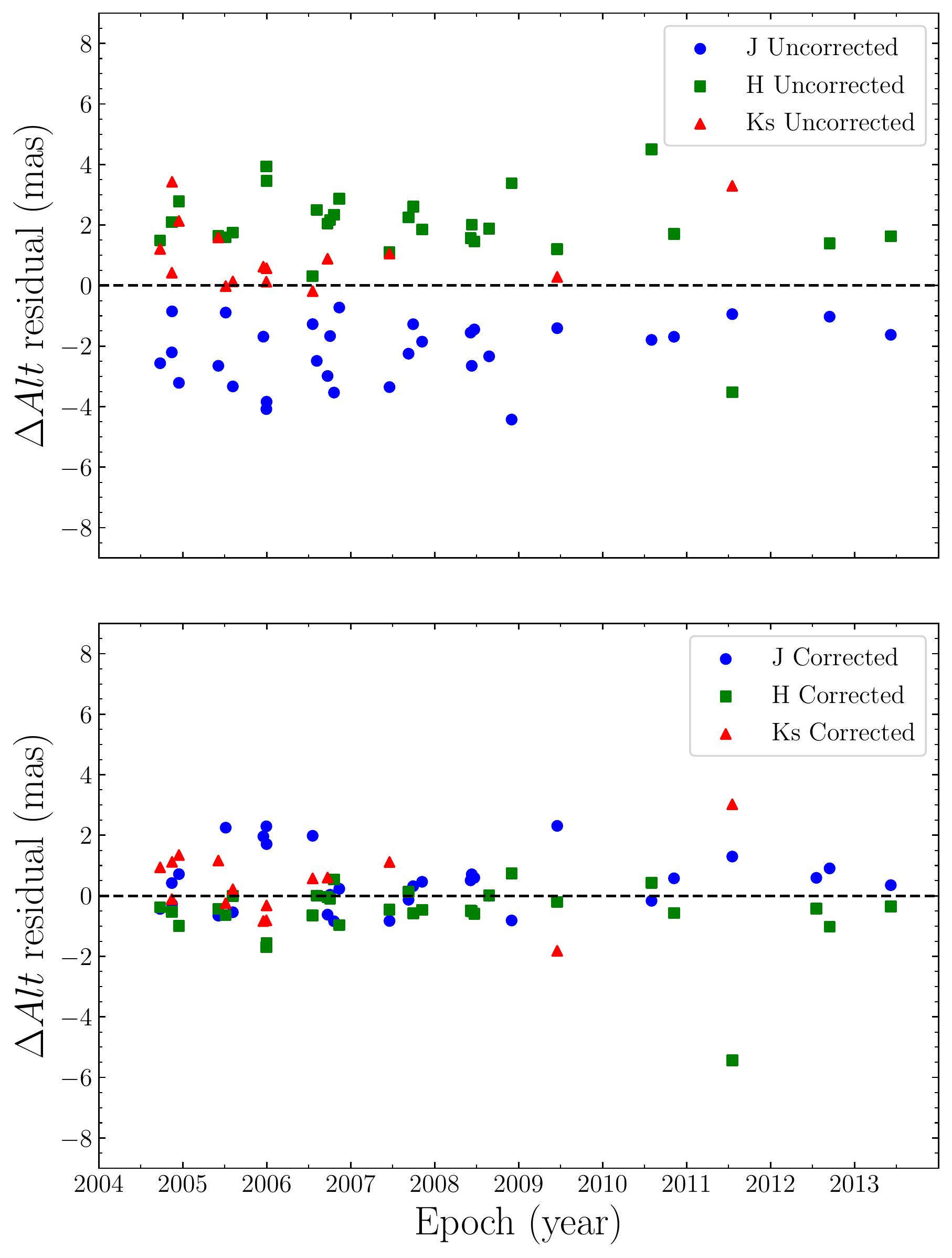}
    \caption{Residual altitude separation of \epsindiba and \Bb in each band compared to the mean of all bands, before (top panel) and after (bottom panel) applying a correction for differential atmospheric refraction.}
    \label{fig:refrac consistency}
\end{figure}

A comparison of the separations along the zenith direction of \epsindiba and \Bb are shown in Figure.\ref{fig:refrac consistency}. We can clearly see the systematic differences between the $J$, $H$, and $K_s$ bands due to differential refraction before the correction. After applying the correction, the three bands are brought to much better agreement as well as having a smaller total scatter around the mean.

{Annual aberration is a phenomenon that describes a change in the apparent position of a light source caused by the observer's changing reference frame due to the orbital motion of the Earth \citep{Bradley_aberration_discovery, Phipps_relativity_and_aberration}. We correct for the differential annual aberration, the difference in aberration between \epsindiba and \Bb, in relative astrometry by transforming the measured positions of \epsindiba and \Bb to a geocentric reference frame using {\tt astropy}. The effect is generally a small fraction of the relative astrometry error bars and has negligible impact on the relative orbit fit. For absolute astrometry, the aberration is absorbed by the linear component of the distortion correction.}

\subsubsection{{PSF Fitting Performance and Systematics}} \label{subsubsec: error inflation}
In order to understand how well our PSF fitting algorithm described in Section \ref{subsec:joint psf fit} performs, we investigate the systematic errors and potential biases of the algorithm in this section, and adjust the errors of our results accordingly. 

To do this, we {crop out boxes around} the stars in the calibration binary, HD~208371/2, {and use them as} empirical PSFs. {We build a collection of such PSF stamps from the images of the calibration binary based on AO quality and SNR. We use these PSFs stamps and empty background regions of the NACO data to generate mock data sets containing overlapping PSFs}. For each such mock image, we randomly select one empirical PSF from the collection and place two copies of this PSF onto the background of an \epsindib image. We scale the fluxes of the two PSF copies to be similar to those of $\varepsilon$ \Ba and \Bb in a typical image. We then generate a large sample of these mock images at various separations and PAs. Since the calibration binary stars are widely separated, these empirical PSFs are effectively free of nearby star contamination. We then perform the PSF fitting described in Section \ref{subsec:joint psf fit} on the mock images and compare the measurements to the true, known separations and PAs.

\begin{figure}
    \centering
    \includegraphics[width=\linewidth]{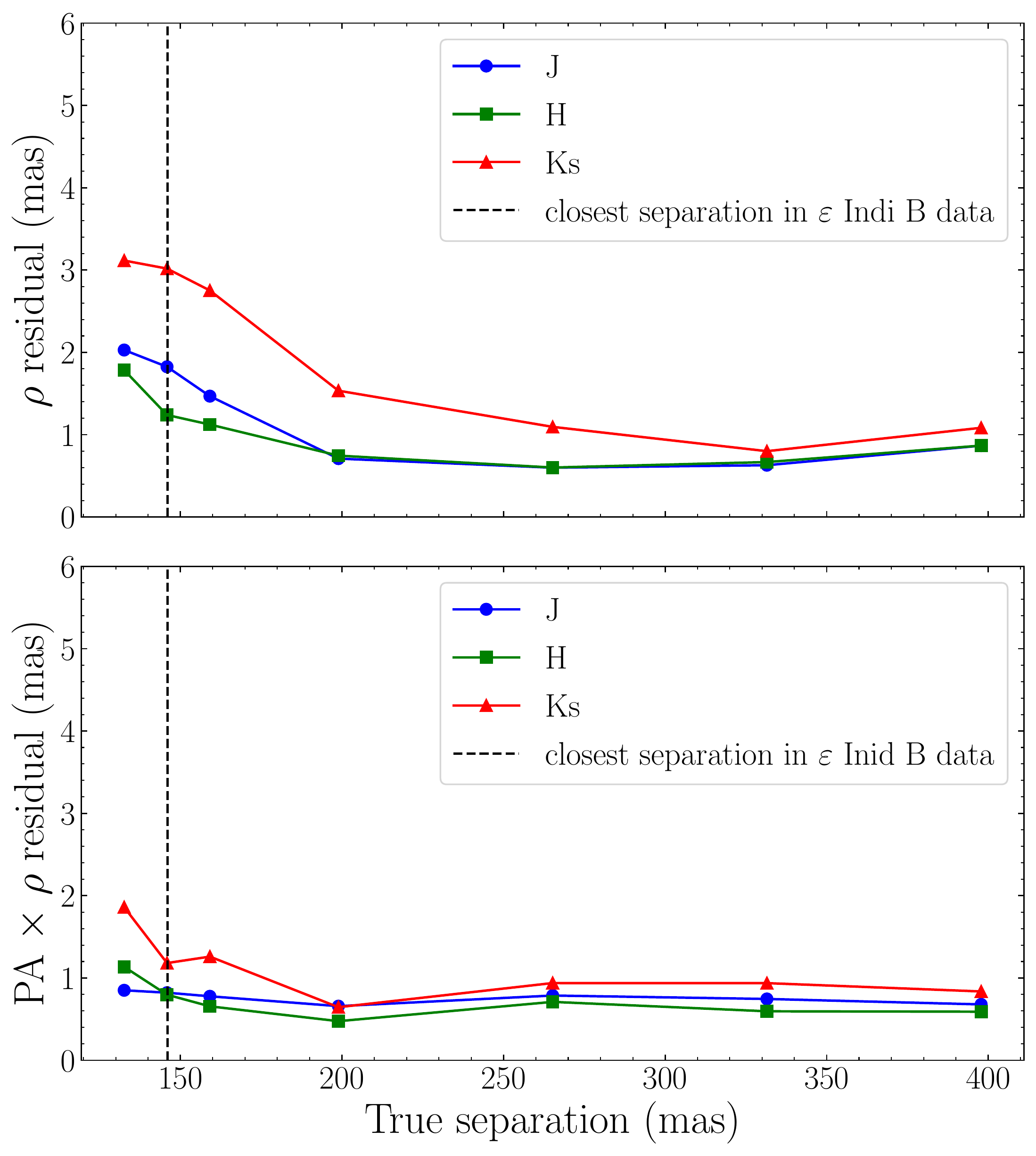}
    \caption{Root mean square residuals of the measured separations from the true separations of the PSFs in simulated data. Top panel shows the residuals in the radial direction. Bottom panel shows the residuals in the tangential direction in terms of arclength. Arclength is a better indicator of the fitting algorithm's performance than PA, because we expect arclength residuals to be independent of radial separation, but the PA will go up at smaller separation simply due to geometry.}
    \label{fig:relast error inflation}
\end{figure}

The results for this test are shown in Figure \ref{fig:relast error inflation}. Each data point is the root mean square residual from fitting 400 mock images at the same separation but with various PAs. The errors we find from these mock data sets are slightly larger but within the same order of magnitude as the scatter in our \epsindib measurements. We also find that the residuals of these mock data measurements do increase as the PSF overlap becomes significant, but they remain at the milliarcsecond level even at a separation equal to the closest separation in the \epsindib data set. The performance for the $K_s$ band is slightly worse due to the large flux ratio of the system in $K_s$. Overall, {our joint PSF fitting algorithm has sub-milliarcsec errors across all three bands for widely separated sources, and has within a few milliarcseconds errors for overlapping sources}. For our final relative astrometry results for \epsindib, we add the systematic errors shown in Figure \ref{fig:relast error inflation} to the measurement errors of the relative astrometry in quadrature.

\subsection{Relative Astrometry Results}

\begin{figure}
    \centering
    \includegraphics[width=\linewidth]{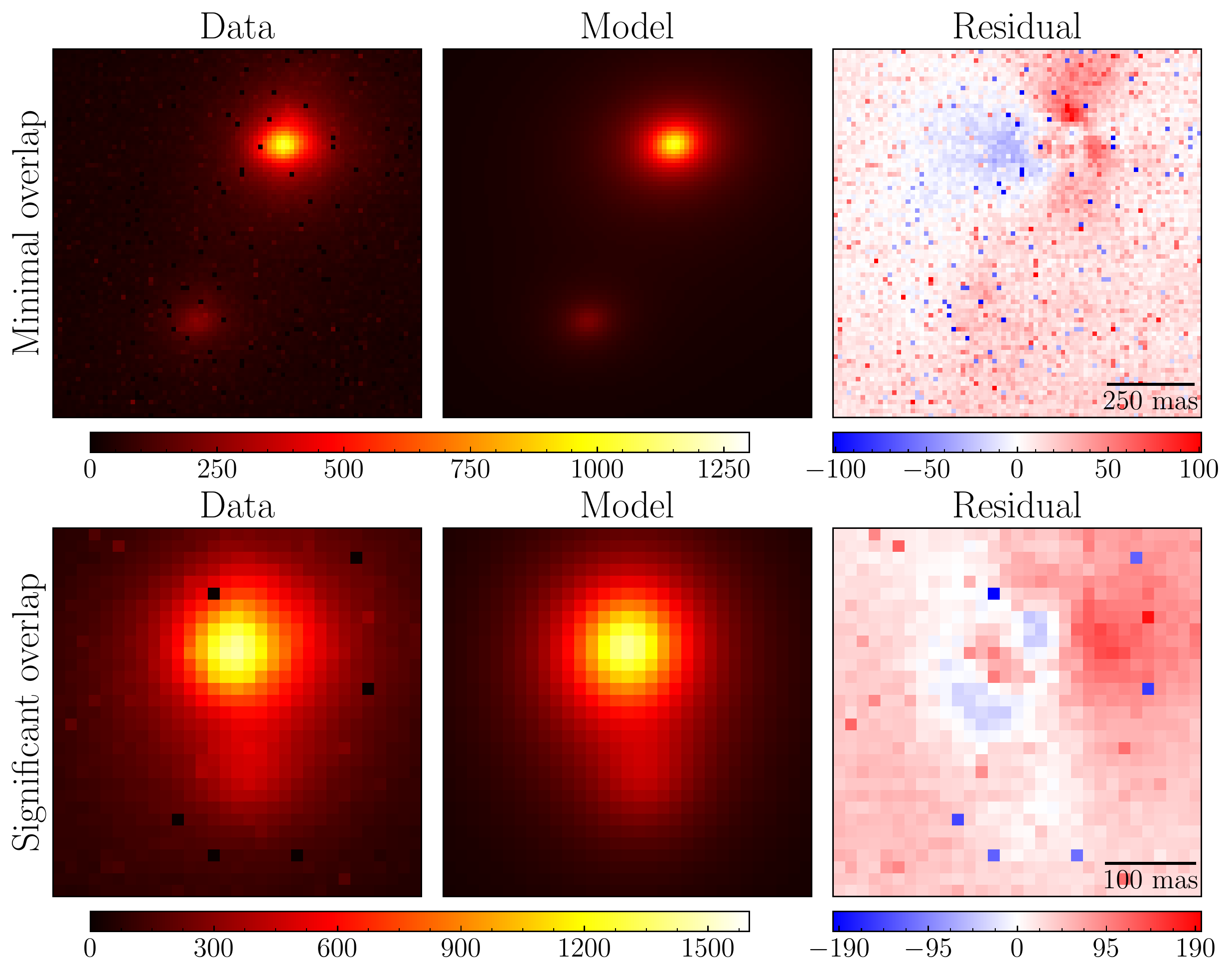}
    \caption{Examples of the joint moffat PSF fits to NACO data. Top panel shows an example of negligible PSF overlap. Bottom panel shows an example from the epoch with maximum overlap in NACO data.}
    \label{fig:relast psf residual}
\end{figure}

The final relative astrometry results are shown in Table \ref{table:relastroresults}. These are measured by applying the calibrations described in Section \ref{subsec: relast calibrations} and jointly fitting for the positions of \epsindiba and \Bb for every selected image, using the PSF fitting method described in Section \ref{subsec:joint psf fit}. We take the mean and the error on the mean for every epoch, and {add the systematic errors quadratically to the measurement errors} as described in Section \ref{subsubsec: error inflation}. In Figure \ref{fig:relast psf residual}, we show a PSF fit for the simple case where the two PSFs are effectively isolated, as well as a PSF fit from the epoch with the closest projected separation and hence maximum PSF overlap. For each case, we take the fit with the median squared residual from that epoch to demonstrate the typical residual level. 

\begin{deluxetable}{ccccc}
\tablecaption{Relative astrometry results \label{table:relastroresults} } 
\tablehead{
\text{Epoch} & \text{$\rho$ (arcsec)} & \text{$\sigma_{\rho}${(arcsec)}} & \text{$\theta$ (deg)} & \text{$\sigma_{\theta}$ {(deg)}}
}
\startdata
2004.730 &  0.88310 &  0.00108 &  140.317 &   0.047 \\
2004.869 &  0.89461 &  0.00110 &  140.853 &   0.047 \\
2004.872 &  0.89560 &  0.00126 &  140.814 &   0.067 \\
2004.954 &  0.90200 &  0.00107 &  141.115 &   0.051 \\
2005.423 &  0.93141 &  0.00112 &  142.648 &   0.045 \\
2005.511 &  0.93351 &  0.00126 &  142.888 &   0.054 \\
2005.595 &  0.93654 &  0.00112 &  143.169 &   0.044 \\
2005.959 &  0.94067 &  0.00118 &  144.222 &   0.050 \\
2005.995 &  0.94079 &  0.00117 &  144.352 &   0.049 \\
2005.997 &  0.93987 &  0.00115 &  144.356 &   0.050 \\
2006.546 &  0.92015 &  0.00111 &  146.044 &   0.053 \\
2006.595 &  0.91721 &  0.00107 &  146.230 &   0.049 \\
2006.724 &  0.90802 &  0.00106 &  146.657 &   0.047 \\
2006.754 &  0.90502 &  0.00110 &  146.745 &   0.047 \\
2006.800 &  0.90222 &  0.00138 &  146.984 &   0.053 \\
2006.863 &  0.89489 &  0.00110 &  147.111 &   0.054 \\
2007.461 &  0.81432 &  0.00109 &  149.295 &   0.050 \\
2007.688 &  0.77183 &  0.00104 &  150.250 &   0.053 \\
2007.743 &  0.76014 &  0.00110 &  150.515 &   0.055 \\
2007.849 &  0.73666 &  0.00109 &  151.017 &   0.063 \\
2008.427 &  0.57619 &  0.00104 &  154.647 &   0.078 \\
2008.441 &  0.57103 &  0.00112 &  154.665 &   0.106 \\
2008.471 &  0.56145 &  0.00110 &  154.978 &   0.086 \\
2008.648 &  0.49830 &  0.00103 &  156.664 &   0.082 \\
2008.915 &  0.39126 &  0.00163 &  160.569 &   0.148 \\
2009.458 &  0.14626 &  0.00273 &  186.175 &   0.562 \\
2010.582 &  0.32838 &  0.00120 &  332.295 &   0.157 \\
2010.849 &  0.30942 &  0.00103 &  339.059 &   0.134 \\
2011.543 &  0.17352 &  0.00243 &   12.950 &   0.433 \\
2012.545 &  0.25518 &  0.00110 &  107.165 &   0.186 \\
2012.703 &  0.29394 &  0.00112 &  112.857 &   0.164 \\
2013.431 &  0.47861 &  0.00104 &  126.845 &   0.088
\enddata
\end{deluxetable}

\subsection{Calibrations for Absolute Astrometry}
\label{sec: absast calibrations}

We now seek to measure the position of \epsindiba relative to a set of reference stars in the FORS2 images with known absolute astrometry. We approach the problem in stages.  First, we fit for the pixel positions of all stars in the frame.  We then use Gaia EDR3 astrometry of a subsample of these stars to construct a distortion map. Next, we use our fit to the NACO data (Section \ref{sec: orbit fit}) to fix the relative positions of \epsindiba and \Bb.  Finally, we use the PSFs of nearby reference stars to model the combined PSF of \epsindiba and \Bb and measure their position in a frame anchored by Gaia EDR3.

We begin by measuring stellar positions in pixel coordinates and using them to derive a conversion between pixel coordinates ($x,y$) and sky coordinates ($\alpha, \delta$), i.e., a distortion correction.  We identify 46 Gaia sources in the field of view of the FORS2 images; these will serve as reference stars to calibrate and derive the distortion corrections. 

We fit elliptical Moffat profiles to retrieve each individual reference star's pixel location $(x, y)$ on the detector. These are Gaia sources with known $\alpha$ and $\delta$ measurements { propagated backwards} from Gaia EDR3's single star astrometry in epoch 2016.0. We adopt the same module used for relative astrometry described in Section~\ref{subsec:joint psf fit} to fit for the reference stars' positions. For each star, we fit for three additional parameters: the FWHMs along two directions, and a rotation angle in between. 

\begin{figure*}
    \centering
    \includegraphics[width=0.8\textwidth]{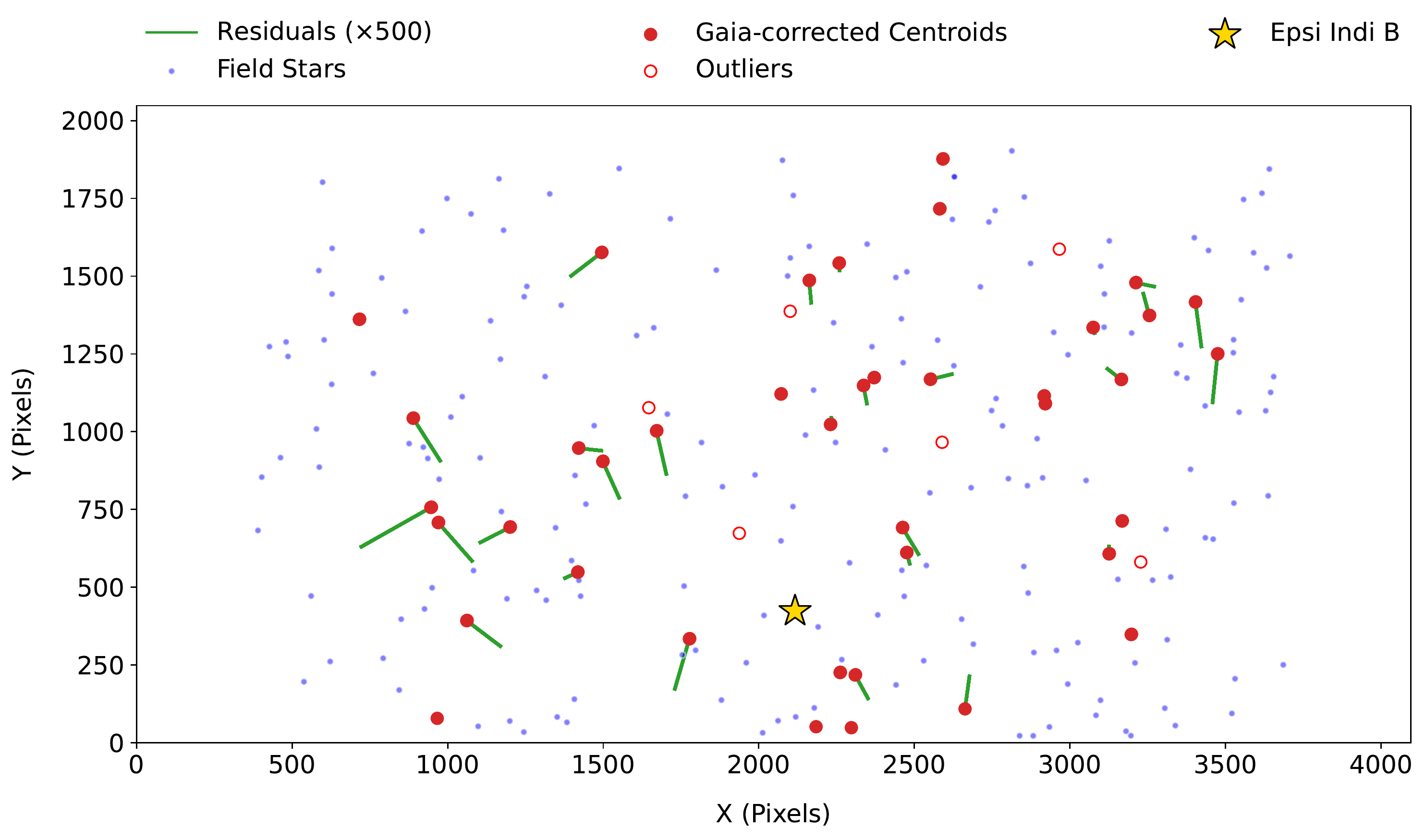}
    \caption{Distortion correction to one of the 904 image frames from the FORS2 long term monitoring of \epsindib's absolute positions. The position of \epsindib is indicated by the yellow star while the blue dots are field stars. The red points are reference stars in the Gaia EDR3 catalog. The green lines indicate the residuals of the measured centroids from their distortion-corrected predictions based on EDR3 astrometry. Red open circles are Gaia EDR3 stars that we discard as outliers. 
    }
    \label{fig::distortion}
\end{figure*}

We assume a polynomial distortion solution of order N for FORS2: 
\begin{align}
    \alpha*_{\rm model} \equiv \alpha \cos \delta &= \sum_{i = 0}^{N} \sum_{j = 0}^{N - i} a_{ij} x^i y^j \\
    \delta_{\rm model} &= \sum_{i = 0}^{N} \sum_{j = 0}^{N - i} b_{ij} x^i y^j .
\end{align}
 
minimizing 

\begin{equation}
    \chi^2 = \sum_{k=1}^{n_{\rm ref}} \left[\left( \frac{{\alpha*_{k}} - {\alpha*_{{\rm model},k}}}{\sigma_{\alpha*,k}} \right)^2 + \left(\frac{{\delta_{k}} - {\delta_{{\rm model},k}}}{\sigma_{\delta_{k}}} \right)^2 \right].
\end{equation}

This defines a linear least-squares problem because each of the $a_{ij}$ appears linearly in the data model. 
To avoid numerical problems, we define $x=y=0$ at the center of the image and subtract $\alpha_{\rm ref} = 181.\!\!^\circ327$, $\delta_{\rm ref} = -56.\!\!^\circ789$ from all Gaia coordinates. { To determine the best model for distortion correction, we compare 2nd, 3rd and 4th-order polynomial models. We derive distortion corrections excluding one Gaia reference star at a time. We then measure the excluded star's positions, and use the distortion correction built without using this star to derive its absolute astrometry.  The consistency of the best-fit astrometric parameters with the Gaia measurements, and the scatter of the individual astrometric measurements about this best-fit sky path, both act as a cross-validation test of the distortion correction.  For most stars, a second-order correction outperforms a third-order correction on both metrics. This also holds true dramatically for \epsindib itself, with a second-order distortion correction providing substantially smaller scatter about the best-fit sky path.} 

Once we have a list of pixel coordinates $(x, y)$ and sky coordinates ($\alpha_{*}$, $\delta$) for all of our reference stars, we derive { second} order distortion corrections for each image. 
To avoid having poorly fit stars drive the results, we clip reference stars that are $\geq$10$\sigma$ outliers. 
Figure~\ref{fig::distortion} shows an example of an image frame indicating the displacement of the distortion-corrected centroids according to Gaia with respect to their original ``uncorrected'' centroid locations on the detector. The empty red circles are Gaia stars that were discarded as outliers. 

We now seek to measure the position of \epsindiba on the distortion-corrected frame defined by the astrometric reference stars.  We cannot fit the brown dwarfs' positions in the same way as the reference stars: their light is blended in most images.  Instead, we first fix their relative position using an orbital fit to the relative astrometry (Sections \ref{subsec: relast calibrations} and \ref{sec: orbit fit}).  We then model the two-dimensional image around \epsindiba and \Bb as a linear combination of the interpolated PSFs of the five nearest field stars.  

With the relative astrometry fixed, our fit to the image around \epsindiba and \Bb has nine free parameters: five for the normalization of each reference PSF, one for the background intensity, one for the flux ratio between \Ba and \Bb, and two for the position of \epsindiba.  The fit is linear in the first six of these parameters.  We solve this linear system for each set of positions and flux ratios, and use nonlinear optimization to find their best-fit values in each image.  We then fix the flux ratio to its median best-fit value of 0.195 and perform the fits again, optimizing the position of \epsindiba in each FORS2 image.

\begin{figure}
    \vskip 0.1 truein
    \includegraphics[width=1\linewidth]{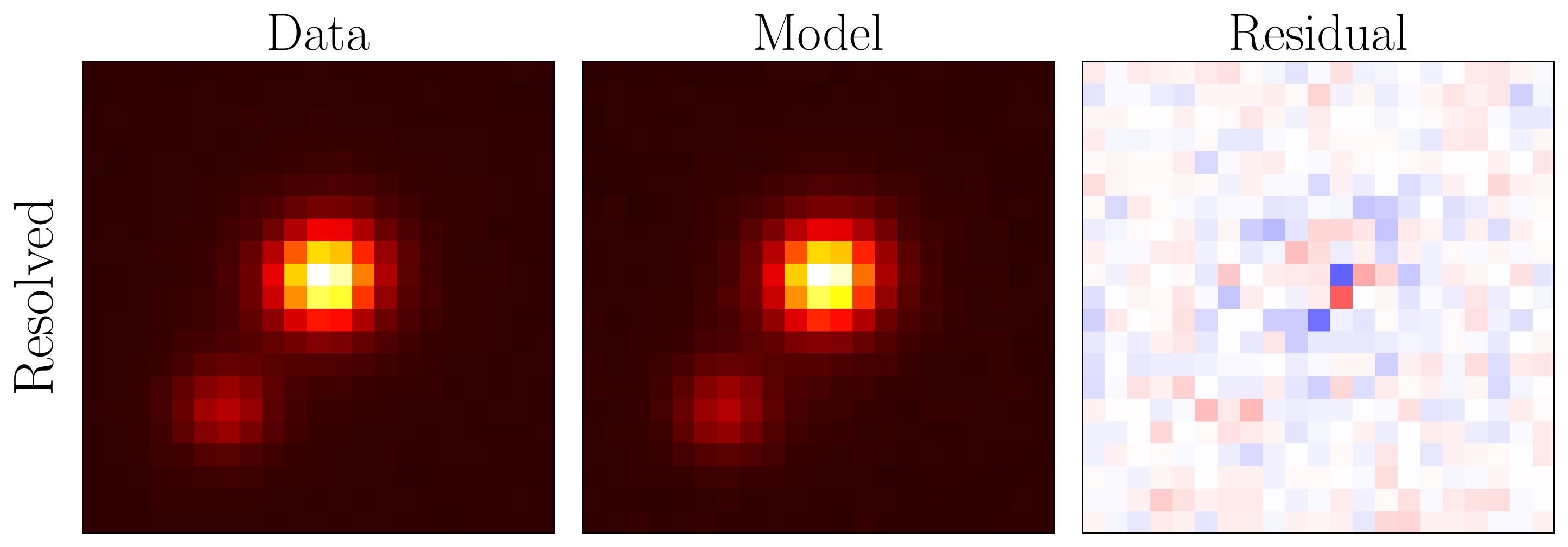} \quad
    \includegraphics[width=1.018\linewidth]{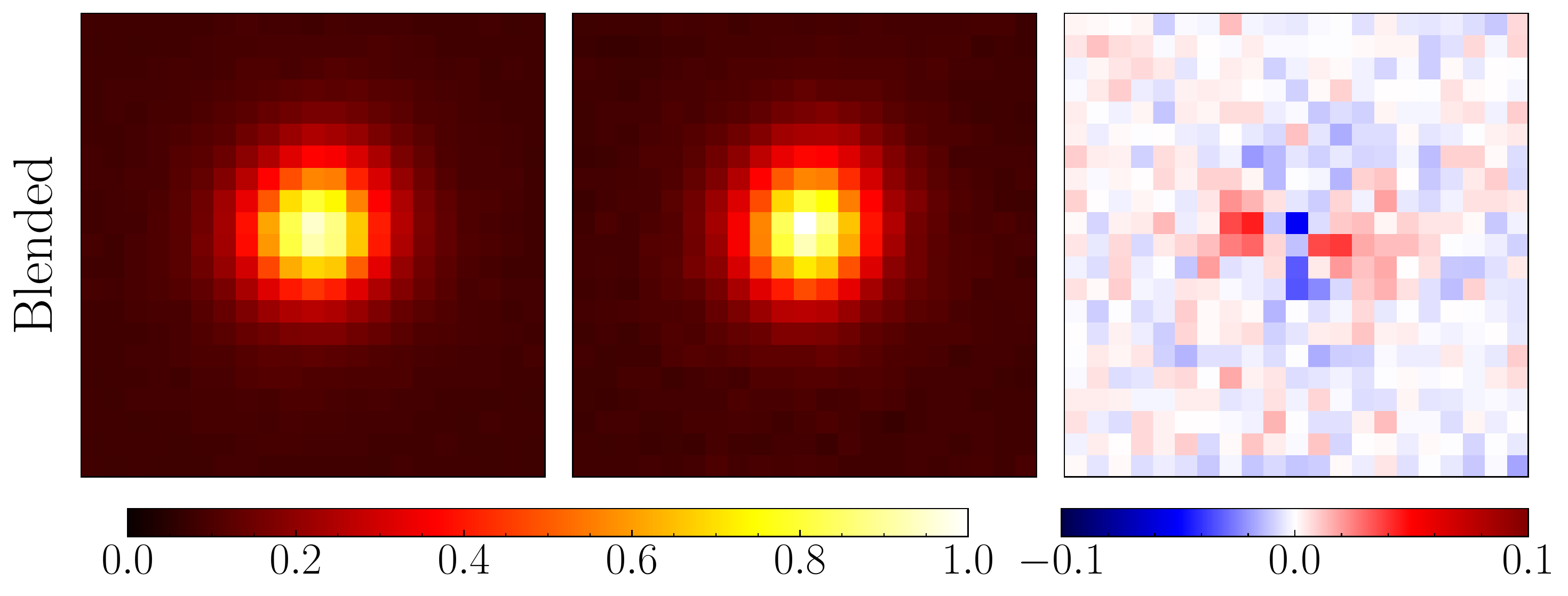}
    \caption{Example fits and residuals to FORS2 images when the two components (\epsindiba and \epsindibb) are completely resolved (top panel) or strongly blended (bottom panel). In each panel, all values are normalized to the peak intensity of the model fits.}
    \label{fig:selfcal_epsiIndib}
\end{figure}

Figure~\ref{fig:selfcal_epsiIndib} shows two examples of the residuals to this fit.  The residual intensity exhibits little structure, whether the two components are strongly blended (bottom panel) or clearly resolved (top panel).  

Our fit produces pixel coordinates of \epsindiba in each frame.  Our use of self-calibration ensures that these pixel coordinates are in the same reference system as the astrometric standard stars.  We then apply the distortion correction derived from these reference stars to convert from pixel coordinates to absolute positions in right ascension and declination. { Another important calibration for absolute astrometry is the correction for atmospheric dispersion. However, our data were taken with an atmospheric dispersion corrector (ADC) in place, which has not been sufficiently well-characterized to model and remove residual dispersion \citep{FORS2_ADC_1997}. We therefore use only the azimuthal projection of the absolute astrometry in the orbital fit. The effects and implications of the ADC and residual atmospheric dispersion are discussed in Section~\ref{sec:adc}. }

\section{Photometric Variability} \label{sec:photvar}

{\cite{Koen_2005}, \cite{Koen_2005_JHK}, and \cite{Koen_2013} found potential evidence of variability of the system in the Near-Infrared ($I$, $J$, $H$ and $K_s$) but also stated that the results are inconclusive due to correlation between seeing and variability.
With the long-term monitoring data acquired by NACO ($J$, $H$, and $K_s$ bands) and FORS2 ($I$ band), we further investigate the photometric variability of \epsindiba and \Bb in this section.}

We apply the generalized Lomb-Scargle method \citep{generalized_Lomb-Scargle_method_2009} and we use the implementation in the {\tt astropy} Python package \citep{astropy:2013, astropy:2018} for this work. For NACO data, there are no other field stars within the FOV to calibrate the photometry. Therefore, we take the best fit flux ratios of \Ba to \Bb from PSF fitting and apply the periodogram on the time series of this flux ratio. For FORS2 data, we use the {\tt photutils} python package to perform differential aperture photometry on the sky-subtracted, flat-fielded and dark-corrected FORS2 images. We first measure the total flux of \epsindiba and \Bb, and the fluxes of fields stars in the field of view. We then normalize the flux of \epsindiba and \Bb using the median flux of all the non-variable field stars to obtain the relative flux of the \epsindi system. We apply the periodogram on this relative flux. We choose a minimum frequency of 0 and a maximum frequency of $1 \; {\rm hour}^{-1}$, which is roughly an upper frequency limit associated with rotational activities if either object were rotating at break-up velocity. We choose a frequency grid size of $\Delta f = 1 / n_0T$, where $n_0 = 10$,  $T=10 \; {\rm yr}$ to sufficiently sample the peaks \citep{VanderPlas_2018}. 

\begin{figure}
    \centering
    \includegraphics[width=1\linewidth]{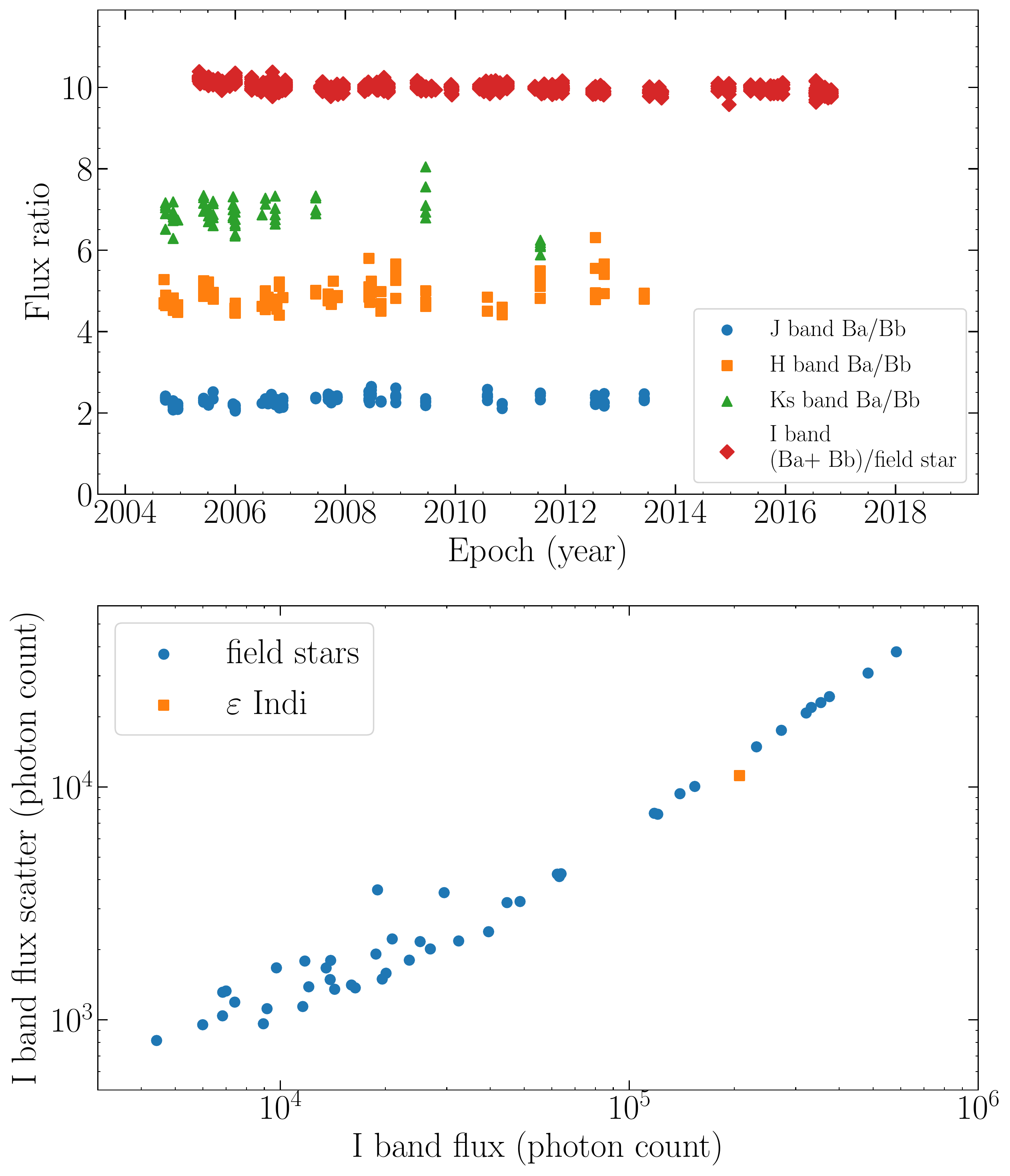}
    \caption{{Top panel shows the flux ratios of \epsindiba over \Bb in J, H and Ks bands measured using the joint PSF fitting method, and the flux ratios of \Ba + \Bb over the average of the field stars measured using aperture photometry. The bottom panel shows the flux scatter of \epsindi compared to the field stars in I band FORS2 data.}}
    \label{fig:light curves}
\end{figure}

{Figure 8 top panel shows the measured flux ratios in both NACO and FORS2 data over all epochs. The bottom panels shows that the \epsindi system has a typical flux scatter for its brightness in FORS2 data. From our simple analysis, we do not see any significant evidence of photometric variability of the system in our periodograms. However, since the observations are not designed for the purpose of investigating variability, the non-uniform and sparsely sampled window function of the observations resulted in very noisy periodograms. Therefore, we also cannot reach any definitive conclusions regarding whether there is any variability of the system with a physical origin.}

\section{Orbital Fit} \label{sec: orbit fit}

\subsection{Relative Astrometry}

\begin{figure*}
    \centering
    \includegraphics[width=\linewidth]{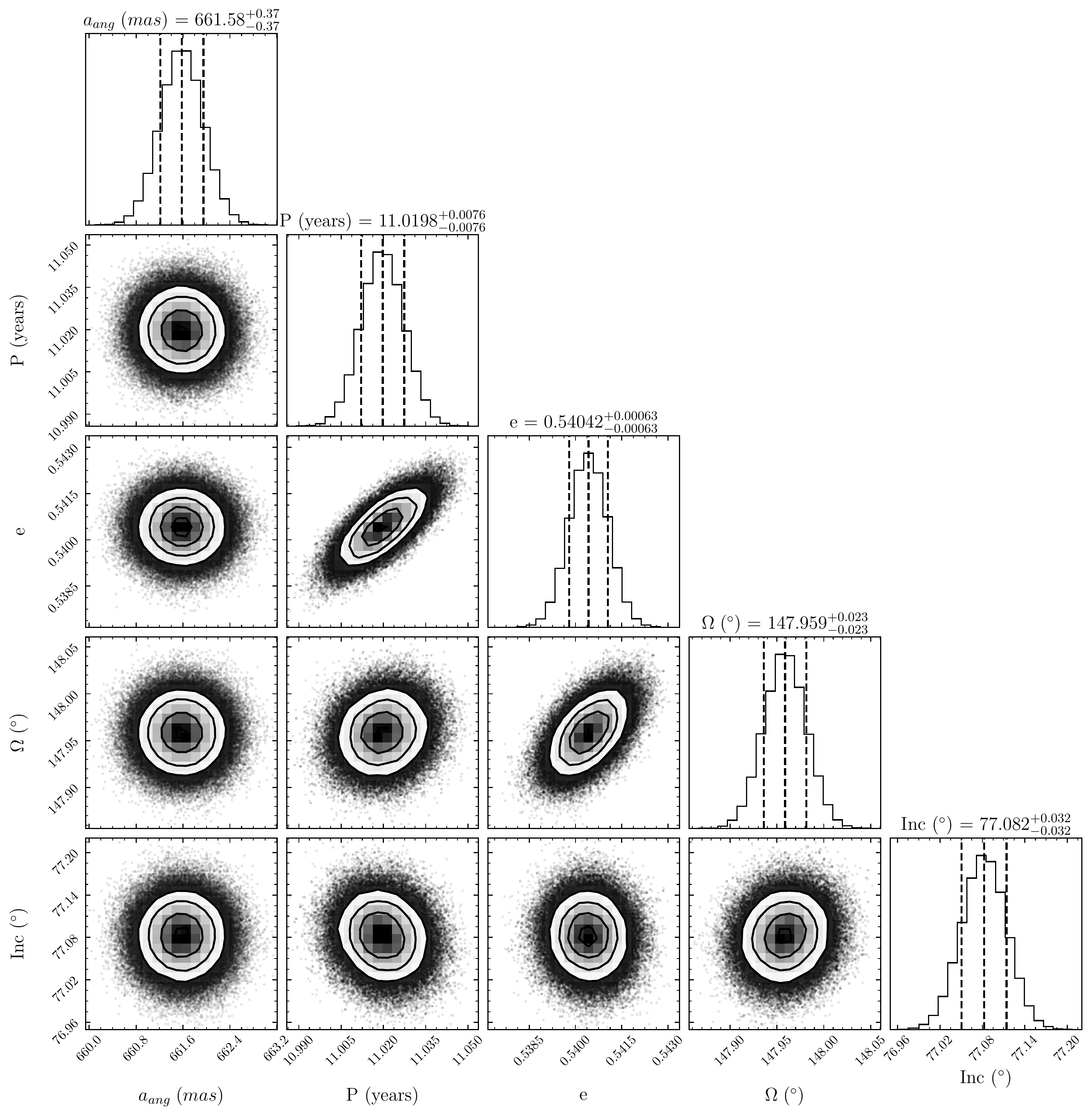}
    \caption{Corner plot for the relative orbit fit MCMC chain. The parameters are angular semi-major axis ($a_{\rm ang}$) in {mas}, period ($P$) in years, eccentricity ($e$), longitude of ascending node ($\Omega$) in degrees and inclination ($i$) in degrees. {The posterior mean is used as the estimator for each parameter, and the errors are one standard deviation from the mean.}}
    \label{fig:rel orbit corner plot}
\end{figure*}

\begin{figure}
    \centering
    \includegraphics[width=\linewidth]{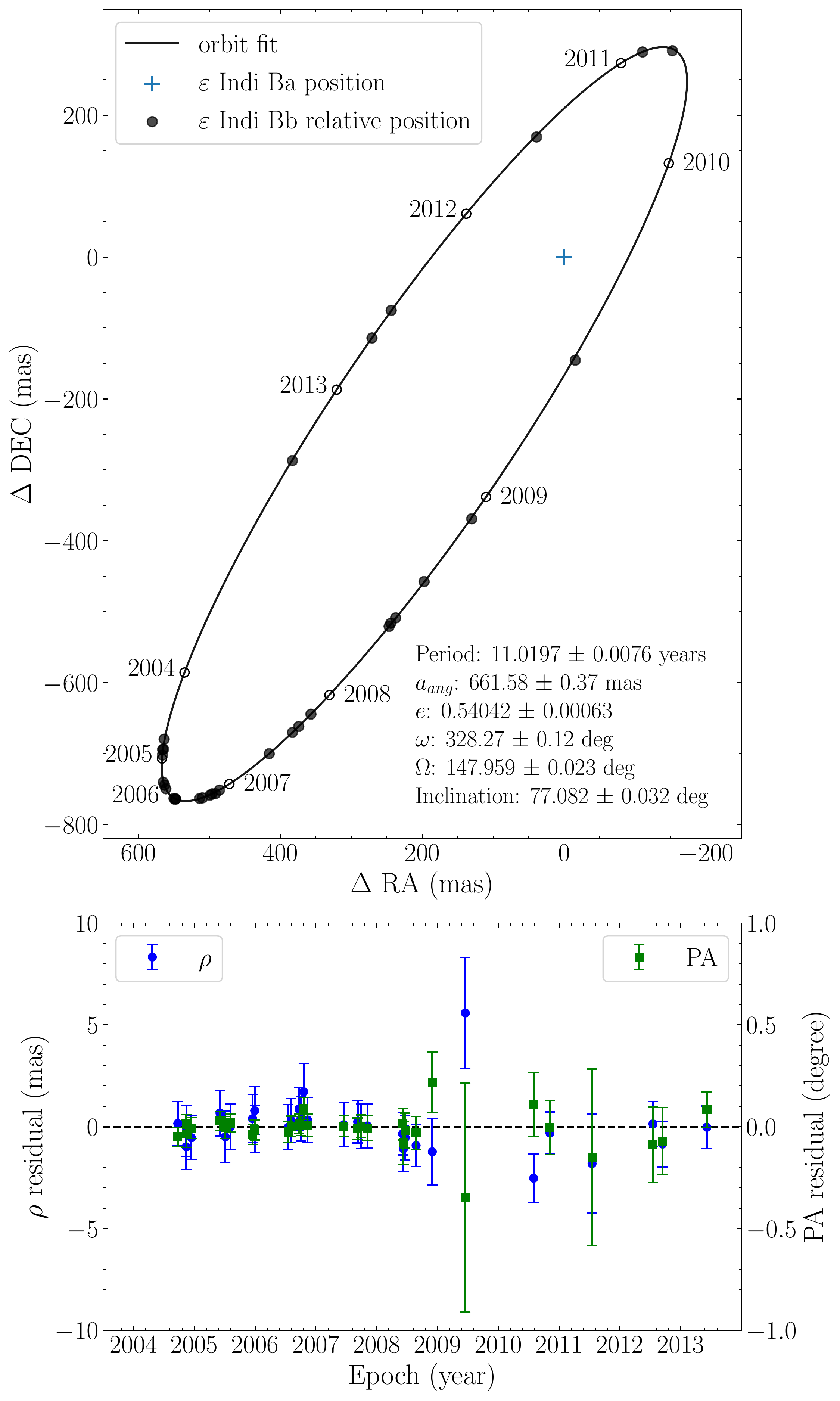}
    \caption{Relative orbit fit of \epsindiba and \Bb. The orbit is plotted as the relative separation of \Bb from \Ba, where \Ba is fixed at the origin. The black dots are the measured relative astrometry, the hollow dots show the beginning of each year, and the solid line is the best fit orbit. The bottom panel shows the residuals of the separation in blue and PA in green.}
    \label{fig:rel orbit}
\end{figure}

We use the relative astrometry measurements summarized in Table \ref{table:relastroresults} to fit for a relative orbit and obtain the orbital parameters. For this, we use an adaptation of the open-source orbital fitting python package, \texttt{orvara} \citep{orvara_2021}, and fit for 7 orbital parameters: period, the angular extent of the semi-major axis ($a_{\rm ang}$, { hereby referred to as the angular semi-major axis}), eccentricity ($e$), argument of periastron ($\omega$), time of periastron ($T_0$), longitude of ascending node ($\Omega$), and inclination ($i$). The corner plot for the MCMC chain is shown in Figure \ref{fig:rel orbit corner plot}. The best fit relative orbit is shown in Figure \ref{fig:rel orbit}, and the best fit orbital parameters are summarized in Table. \ref{table:orbitfit}. The reduced $\chi^2$ is 0.77 which suggests that we may be slightly overestimating the errors, especially for the earlier epochs. This is possibly because the earlier epochs in general have higher quality data, while we used empirical PSFs of a wider range of qualities in order to generate a large enough sample for the error inflation estimate described in section \ref{subsubsec: error inflation}. Nevertheless, we are able to produce an excellent fit and obtain very tight constraints on the orbital parameters thanks to high quality direct imaging data and a long monitoring baseline that almost covers an entire period.

\subsection{Absolute Astrometry}

We have derived optical geometric distortion corrections for all the FORS2 images in Section~\ref{sec: absast calibrations}. We describe here our approach to fit for astrometric models to the reference stars, field stars, and most importantly the \epsindib system. We fit standard five-parameter astrometric models, with position, proper motion, and parallax, to the reference stars and field stars in the field of view of the FORS2 images. The results from the fits for reference stars match within 20$\%$ from the proper motions and parallaxes Gaia provided.
For the binary system \epsindib, we fit a six-parameter astrometric solution, adding an extra parameter which is the ratio between the semi-major axes of the orbits of the two components. { We also review, and ultimately project out, the effects of atmospheric dispersion. The wavelength-dependent index of refraction of air causes an apparent, airmass-dependent displacement between the redder brown dwarfs \epsindiba and \Bb and the bluer field stars along the zenith direction.}

The results from absolute astrometry give proper motions, parallax, and a ratio between the semi-major axes which can then be converted into a mass ratio and individual masses. In conjuction with our previous relative astrometry results, full Keplerian solutions can be derived that completely characterize the orbits of both \epsindiba and \epsindibb. 

\subsubsection{Astrometric Solution}

The astrometric solution for a single and isolated reference star or background star is a five-parameter linear model in terms of reference pixel coordinates in RA and Declination, proper motions in RA and Declination, and the parallax. A star's instantaneous position $(\alpha*, \delta)$ would be its position $(\alpha*_{\rm ref}, \delta_{\rm ref})$
at a reference epoch, plus proper motion $(\mu_{\alpha*}, \mu_\delta)$ multiplied by the time since { the reference epoch $ t_{\rm ref}$}, and parallax $\varpi$ times the so-called parallax factors $\Delta \pi_{\alpha*}$ and $\Delta \pi_\delta$:

\begin{equation}
\begin{bmatrix}
1 & 0 & t - t_{\rm ref} & 0 & \rm \Delta \pi_{\alpha*} \\
0 & 1 & 0 & t - t_{\rm ref} & \rm \Delta \pi_{\delta} \\
\end{bmatrix} 
\begin{bmatrix}
{\alpha*_{\rm ref}} \\ \delta_{\rm ref} \\
\mu_{\alpha*} \\
\mu_{\delta} \\
\varpi
\end{bmatrix}
=
\begin{bmatrix}
\alpha* \\
\delta \\
\end{bmatrix}.
\end{equation}

To test the robustness of the distortion corrections in RA and Declination that we have derived for each image, we `reverse engineer' by excluding a particular reference star from the fit and solve for the astrometric solution of that star based on the discussion above for comparison to the Gaia parameters. In particular, we focus on the reference stars close to \epsindib. 

For the binary system \epsindib, the astrometric solution demands an additional parameter $r_{\rm Ba}$: the ratio between the semi-major axis of \epsindiba about the barycenter to the total semi-major axis $a$. The parameter $r_{\rm Ba}$ is related to the binary mass ratio by 
\begin{equation}
    r_{\rm Ba} = \frac{M_{\rm Bb}}{M_{\rm Ba} + M_{\rm Bb}}.
\end{equation}
The model becomes
\begin{equation}
\begin{split}
\begin{bmatrix}
1 & 0 & t - t_{\rm ref} & 0 & \rm \Delta \pi_{\alpha*} &  a_{\alpha*}\\
0 & 1 & 0 & t - t_{\rm ref} & \rm \Delta \pi_{\delta} & a_{\delta}\\
\end{bmatrix} 
\begin{bmatrix}
\alpha*_{\rm ref} \\
\delta_{\rm ref} \\
\mu_{\alpha*} \\
\mu_{\delta} \\
\varpi \\
r_{\rm Ba}
\end{bmatrix}
\\
=
\begin{bmatrix}
\alpha \!*_{\rm Ba}-0.5\dot{\mu}_{\alpha*} (t-t_{\rm ref})^2 \\
\delta_{\rm Ba}- 0.5 \dot{\mu}_{\delta}(t-t_{\rm ref})^2 \\
\end{bmatrix}.
\label{eq:absast_fit}
\end{split}
\end{equation}

{ We also take into account the perspective acceleration that occurs when a star passes by the observer and its proper motion gets exchanged into radial velocity. This effect is more significant for \epsindi than for remote stars. We employ constant perspective accelerations of $\dot{\mu}_{\alpha*}$ = 0.165 mas\,yr$^{-2}$ in RA and $\dot{\mu}_{\delta}$ = 0.078 mas\,yr$^{-2}$ in Dec for the \epsindib system based on Gaia EDR3 measurements and assuming the radial velocity measured for \epsindi\,A. We adopt a reference epoch $t_{\rm ref}$=2010. With an astrometric baseline of $\sim$10 years, this gives a displacement of $0.5\dot{\mu}(t-t_{\rm ref})^2\approx 2$\,mas at the edges of the observing window, where $\dot{\mu}$ is the acceleration. The perspective acceleration, because it is known, is included in the right hand side of Equation \eqref{eq:absast_fit}.}

\begin{figure*}
    \centering
    \includegraphics[width=0.8\textwidth]{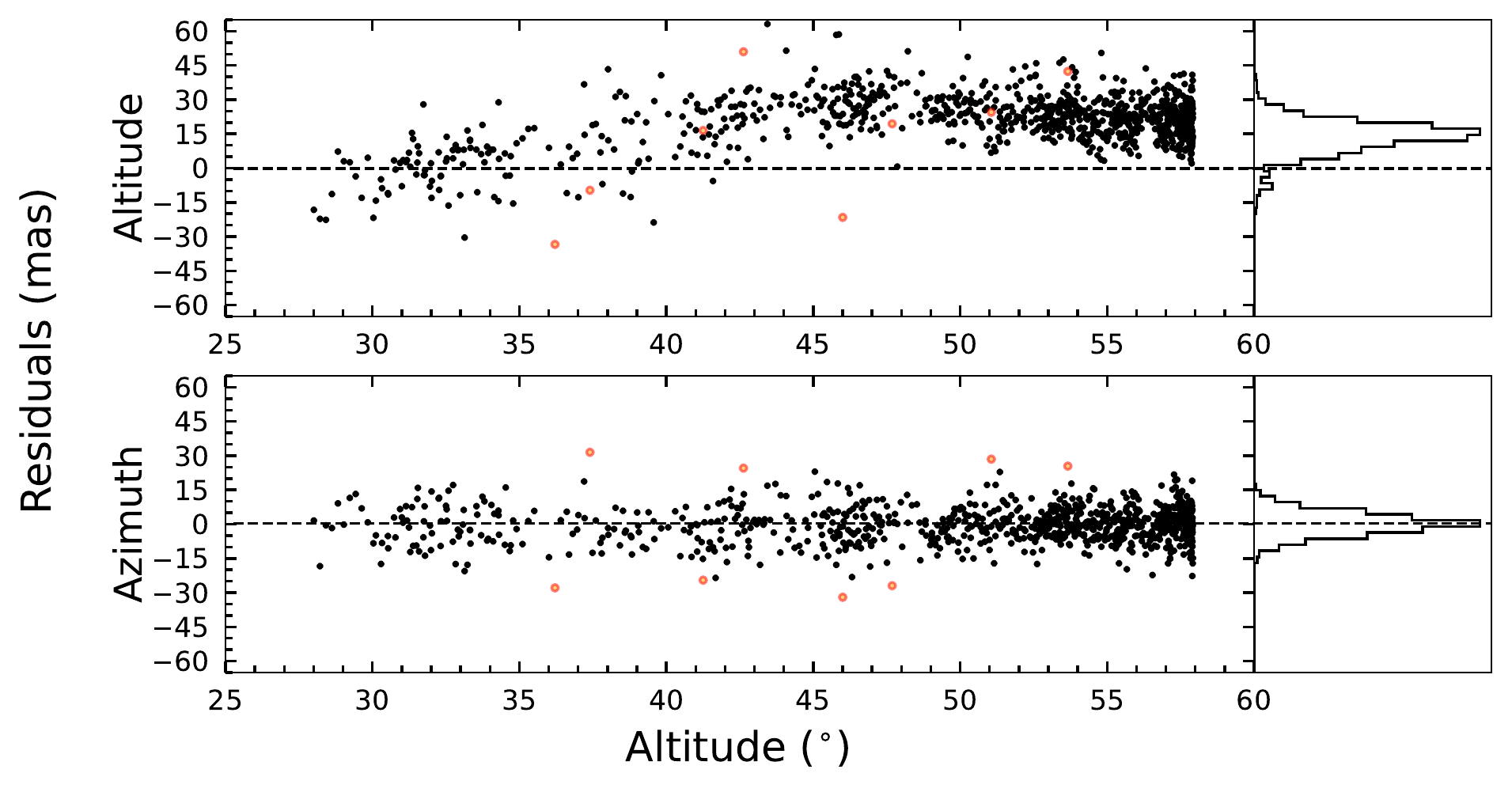}
    \caption{Residuals to the best-fit astrometric model for \epsindiba. The top panel shows the residuals in altitude, and the bottom panel shows the residuals in azimuth, both plotted as a function of altitude. Empty red circles show the rejected epochs from 3$\sigma$ clipping of the azimuthal residuals. A histogram of the residuals is shown to the right of each scatter plot. Strong systematics are seen in the altitude residuals but not the azimuth ones, evident from the symmetric, roughly Gaussian distribution for the former and the altitude-dependent nonzero mean for the latter. These systematics are consistent with the magnitude expected for uncorrected ADC residual dispersion \citep{FORS2_ADC_1997}. \label{fig:residuals}}
\end{figure*}

\subsubsection{ Residual Atmospheric Dispersion}
\label{sec:adc}

The FORS2 imaging covers twelve years, with data taken over a wide range of airmasses. This makes it essential to correct for atmospheric dispersion caused by the differential refraction of light of different colors as it passes through the atmosphere. The degree of dispersion is related to the wavelength of light, the filter used, and the airmass, but is always along the zenith direction. 
Many of the FORS2 images were taken at very high airmass. The typical airmass will vary over the course of the year, because the time of observation will vary depending on what part of night the target is up.

All of the FORS2 images were taken with an atmospheric dispersion corrector, or ADC.  The residual dispersion depends on filter, airmass, and position on the FOV, but is typically tens of mas \citep{FORS2_ADC_1997}.  This is smaller than the system's parallax and { angular} semi-major axis, but only by a factor of $\approx$10.  { Further, the ADC is only intended to provide a full correction to a zenith angle of $50^\circ$ \citep{FORS2_ADC_1997}.  At lower elevations it is parked at its maximum extent; \cite{Cardoso_2012} applied an additional correction to these data.} Because the { residual dispersion is} only in the zenith direction, we perform two fits to the absolute astrometry of \epsindiba.  First, we use our measurements in RA and Decl.~directly.  Second, we use the parallactic angle $\theta$ to take only the component of our measurement along the azimuth direction, which is immune to the effects of differential atmospheric refraction.

We project the data into the altitude-azimuth frame by left-multiplying both sides of Equation \eqref{eq:absast_fit} by the rotation matrix
\begin{equation}
    { R} = 
    \begin{bmatrix}
    \cos \theta & -\sin \theta \\
    \sin \theta & \cos \theta
    \end{bmatrix}.
    \label{eq:rot_altaz}
\end{equation}
The top row of Equation \eqref{eq:rot_altaz} corresponds to the azimuth direction, while the bottom row corresponds to the altitude direction.

Fitting in both the altitude and azimuth directions produces a parallax of 263~mas, in agreement with \cite{Cardoso_2012} but much lower than both the Hipparcos and Gaia values for \epsindi~A.  We then perform a fit only in the azimuth direction: we multiply both sides of Equation \eqref{eq:absast_fit} by the top row of Equation \eqref{eq:rot_altaz}.  
We exclude eight $3\sigma$ outliers and assume a uniform per-epoch uncertainty of 8.01 mas to give a normalization factor that gives a reduced $\chi^2$ value of 1.00. This procedure results in a parallax of $274.99 \pm 0.43$\,mas, in good agreement with both the Hipparcos and Gaia measurements. { We note that the 25-year time baseline between Hipparcos and Gaia causes a small parallax difference. \epsindi~A has a radial velocity of 40.5 km/s \citep{GaiaDR2_2018}, which translates to a fractional change of $3\times 10^{-4}$ or a decrease in parallax of about 0.08 mas over 25 years. This difference is much smaller than the uncertainties of any of these parallax measurements.}

Figure~\ref{fig:residuals} shows the residual to the best-fit model using only azimuthal measurements: the top panel shows the residuals in altitude, while the bottom panel shows the residuals in azimuth. A upward trend and nonzero mean are seen in the altitude component of the parallax as a function of altitude, but no dependence on altitude was seen in the azimuth-based parallax. This confirms that the altitude component of the position measurements is corrupted by residual atmospheric dispersion of a magnitude consistent with expectations \citep{FORS2_ADC_1997}. 

The six-parameter azimuth-component-only astrometric solution gives a mass ratio of 0.4431 $\pm$ 0.0008 between the binary brown dwarf \epsindiba and \Bb. This mass ratio is consistent with \cite{Cardoso_2012}. The only differences in our approaches arise from our usage of Gaia EDR3 to anchor the distortion correction and our account of atmospheric dispersion by only taking the azimuthal projection of the motion of the system. Our parallax of $274.99 \pm 0.43$ mas agrees with the parallax value from Hipparcos and that of \citet{Dieterich_2018}. 

\subsection{Individual Dynamical Masses} 

The relative astrometry orbital fit provides a precise period and angular semi-major axis.  With a parallax from absolute astrometry, we convert the angular semi-major axis to distance units: $2.4058 \pm 0.0040$\,au.  We then use Kepler's third law to calculate a total system mass of \MB\,$M_{\rm Jup}$.  Finally, the mass ratio derived from absolute astrometry provides individual dynamical masses of \MBa\ and \MBb\,$M_{\rm Jup}$ for \epsindiba and \Bb, respectively.  Table \ref{table:orbitfit} shows the results of each component of the orbital fit, with the final individual mass measurements in the bottom panel.  The uncertainty on these masses is dominated by uncertainty in the parallax: the mass ratio is constrained significantly better than the total mass.  In the following section, we use both our individual mass constraints and our measurement of the mass ratio to test models of substellar evolution.

\begin{deluxetable}{cc}
\tablewidth{0pt}
    \tablecaption{Orbital Fit of the \epsindib system \label{table:orbitfit}}
    \tablehead{
    \colhead{{ Fitted Parameters}} & \colhead{Posterior {mean} $\pm$1$\sigma$}}
    \startdata
    Period (yr) & $11.0197 \pm 0.0076\phn$ \\
    { Angular} semi-major axis (mas) & $661.58 \pm 0.37\phn\phn$ \\
    Eccentricity & $0.54042 \pm 0.00063$ \\
    $\omega$ (deg) & $328.27 \pm 0.12\phn\phn$ \\
    $\Omega$ (deg) & $147.959 \pm 0.023\phn\phn$ \\
    Inclination (deg) & $77.082 \pm 0.032\phn$ \\
    $\mu_{\alpha*}$ (\masyr) & $3987.41 \pm 0.12 \phn\phn\phn$ \\
    $\mu_{\delta}$ (\masyr) & $-2505.35 \pm 0.10\phn\phn\phn$\phs \\
    $\Big(\frac{M_{\rm Bb}}{M_{\rm Ba}+M_{\rm Bb}}\Big)$ & $0.4431 \pm 0.0008$ \\
    $\varpi$ (mas) & $274.99 \pm 0.43\phn\phn$\\
    reduced $\chi^2$ & 1.00\\
    \hline
    { Derived Parameters} & Posterior {mean} $\pm$1$\sigma$\\
    \hline
    a (AU) & $2.4058 \pm 0.0040$ \\
    System mass ($\Mjup$) &  \MB$\phn\phn$ \\
    ${\rm Mass}_{\rm Ba}$ ($\Mjup$) & \MBa$\phn$\\
    ${\rm Mass}_{\rm Bb}$ ($\Mjup$) & \MBb$\phn$
    \enddata
\end{deluxetable}

\section{Testing Models of Substellar Evolution} \label{sec:BDtests}

The evolution of substellar objects is characterized by continuously-changing observable properties over their entire lifetimes. Therefore, the most powerful tests to benchmark evolutionary models utilize dynamical mass measurements of brown dwarfs of known age (usually, from an age-dated stellar companion) or of binary brown dwarfs that can conservatively presumed to be coeval. 
A single brown dwarf of known age and mass can test evolutionary models in an absolute sense, and the strength of the test is limited by both the accuracy of the age and of the mass. Pairs of brown dwarfs of known masses can test the slopes of evolutionary model isochrones, even without absolute ages, because their age difference is known very precisely to be near zero unless they are very young. 

The \epsindib system is an especially rare case where both of these types of tests are possible. In fact, it is the only such system containing T~dwarfs where both the absolute test of substellar cooling with time and coevality test of model isochrones are possible.

In the following, we consider a collection of evolutionary models applicable to \epsindiba and \Bb covering a range of input physics. The ATMO-2020 grid \citep{2020A&A...637A..38P} represents the most up-to-date cloudless evolutionary models from the ``Lyon'' lineage that includes DUSTY \citep{Chabrier:2000sh}, COND \citep{2003A&A...402..701B}, and BHAC15 \citep{2015A&A...577A..42B}. For models that include the effect of clouds, we use the hybrid tracks of \citet[][hereinafter SM08]{2008ApJ...689.1327S}, which are cloudy at $\Teff>1400$\,K, cloudless at $\Teff<1200$\,K, and a hybrid of the two in between 1400\,K and 1200\,K. These are the most recent models that include cloud opacity from the ``Tucson'' lineage. We also compare to the earlier cloudless Tucson models \citep{Burrows1997} given their ubiquity in the literature.

In order to test these models, we chose pairs of observable parameters from among the fundamental properties of mass, age, and luminosity. Using any two parameters, we computed the third from evolutionary models. When the first two parameters were mass and age, we bilinearly interpolated the evolutionary model grid to compute luminosity. When the first two parameters were luminosity and either mass or age, we used a Monte Carlo rejection sampling approach as in our past work \citep{Dupuy+Liu_2017,2018AJ....156...57D,Brandt2021_Six_Masses}. Briefly, we randomly drew values for the observed independent variable, according to the measured mass or age posterior distribution, and then drew values for the other from an uninformed prior distribution (either log-flat in mass or linear-flat in age). We then bilinearly interpolated luminosities from the randomly drawn mass and age distributions. For each interpolated luminosity $L_{\rm bol}^{\prime}$, we computed $\chi^2 = (\Lbol-L_{\rm bol}^{\prime})^2/\sigma_{L_{\rm bol}}^2$. For each trial we drew a random number between zero and one, and we only retained trial sets of mass, age, and luminosity in our output posterior if $e^{-(\chi^2-\chi^2_{\rm min})/2}$ was greater than the random number.

We used the luminosities of \epsindiba and \Bb from \citet{2010A&A...510A..99K}, accounting for the small difference between the \Hipparcos\ parallax of 276.06\,mas that they used and our value of 274.99\,mas, which resulted in $\log(\Lbol/\Lsun) = -4.691\pm0.017$\,dex and $-5.224\pm0.020$\,dex. Their luminosity errors were dominated by their measured photometry of \epsindiba and \epsindibb and the absolute flux calibration of Vega's spectrum, so our errors are identical to theirs.

Our Monte Carlo approach naturally accounts for the relevant covariances between measured parameters. There are six independently-measured parameters for which we randomly drew Gaussian-distributed values: the orbital period ($P$), the semi-major axis in angular units ($a^{\prime\prime}$), the ratio of the mass of \epsindibb to the total mass of \epsindib ($M_{\rm Bb}/M_{\rm tot}$), the parallax in the same angular units as the semi-major axis ($\varpi$), and the two bolometric fluxes computed from the luminosities and distance in \citet{2010A&A...510A..99K}. From these, we computed the total mass, $M_{\rm tot} = (a^{\prime\prime}/\varpi)^3 (P/1{\rm yr})^{-2}$, and the individual masses and luminosities.

\begin{figure*}
    \centering
    \includegraphics[width=0.32\textwidth]{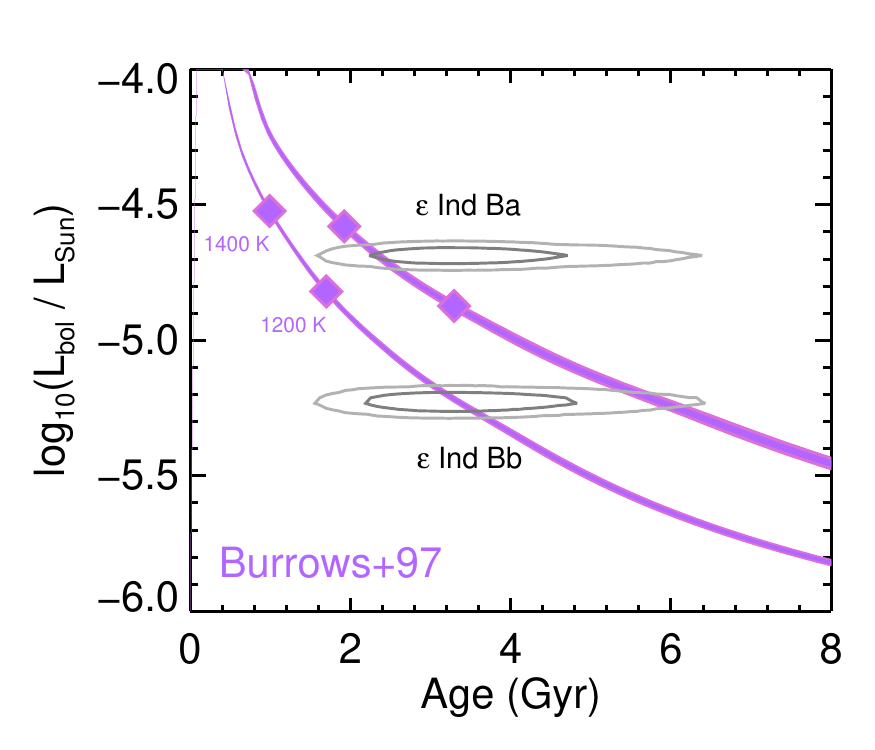}
    \includegraphics[width=0.32\textwidth]{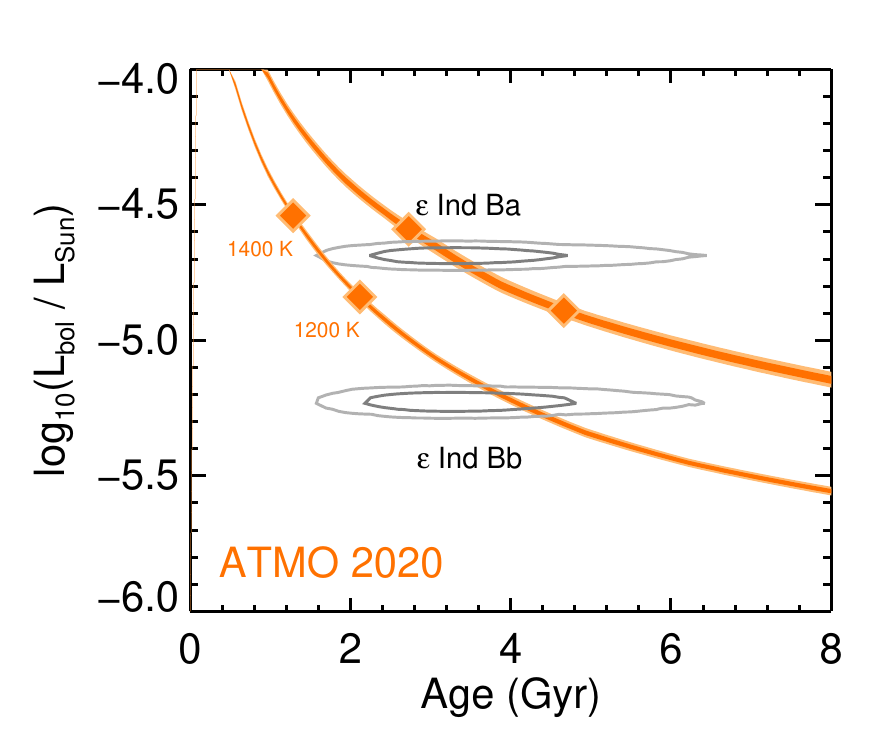}
    \includegraphics[width=0.32\textwidth]{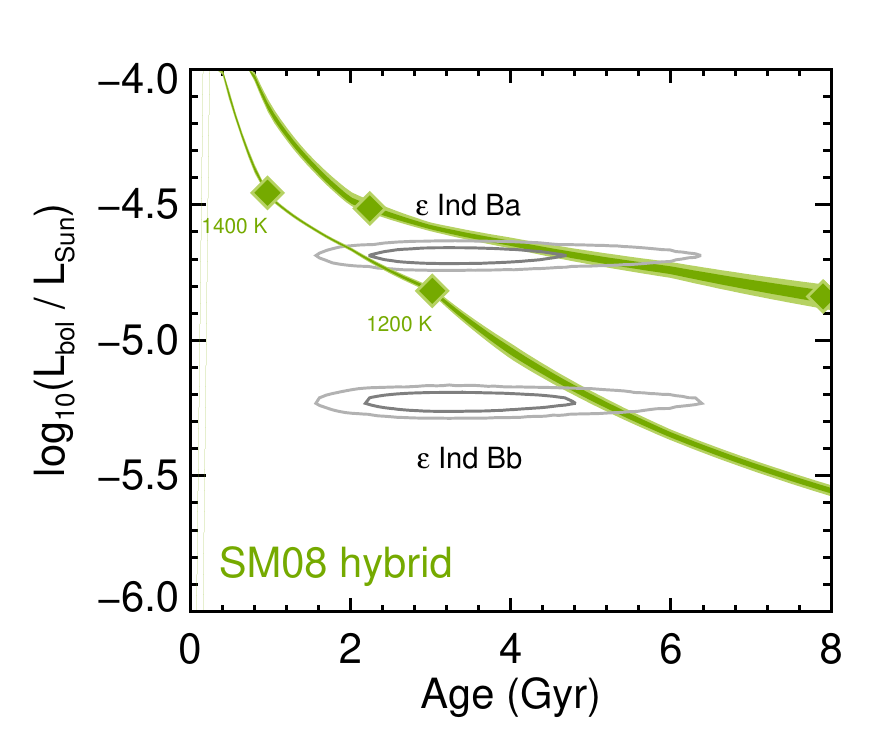}
    \caption{Substellar cooling curves derived from three independent evolutionary models given our measured masses. The top curve in each panel corresponds to \epsindiba, and the bottom curve corresponds to \epsindibb. The darker shaded region of each curve shows the 1$\sigma$ range in our measured mass, and the lighter shading is the 2$\sigma$ range. On each curve, the ages corresponding to $\Teff = 1400$\,K and 1200\,K are marked, indicating the approximate beginning and ending of the L/T transition. Over-plotted on each panel are the 1$\sigma$ and 2$\sigma$ joint uncertainty contours for the age and luminosities of \epsindiba and \Bb.}
    \label{fig:lbol-age}
\end{figure*}

\subsection{Absolute test of $\Lbol(t)$}

In general, tests of substellar luminosity as a function of time are either dominated by the uncertainty in the age or in the mass. In the case of the \epsindib system, with highly precise masses having 0.5\% errors, the uncertainty in the system age ($t = 3.5^{+0.8}_{-1.0}$\,Gyr) is by far the dominant source of uncertainty. 

Figure~\ref{fig:lbol-age} shows the measured joint confidence intervals on luminosities and age of \epsindiba and \Bb compared to evolutionary model predictions given their measured masses. The measured luminosity-age contours overlap all model predictions to within $\approx$1$\sigma$ or less for both components. To quantitatively test models and observations, we compared our model-derived substellar cooling ages for \epsindiba and \Bb to \epsindi~A's age posterior, finding that they are all statistically consistent with the stellar age.

Our results for \epsindiba and \Bb are comparable to other relatively massive (50--75\,\Mjup) brown dwarfs of intermediate age (1--5\,Gyr) that also broadly agree with evolutionary model predictions of luminosity as a function of age \citep{Brandt2021_Six_Masses}. These include objects such as HR~7672~B \citep{Brandt_2019}, HD~4747~B \citep{Crepp+Principe+Wolff+etal_2018}, HD~72946~B \citep{Maire+Baudino+Desidera_2020}, and HD~33632~Ab \citep{2020ApJ...904L..25C}.

However, despite agreeing with models in an absolute sense, it is evident in Figure~\ref{fig:lbol-age} that the ATMO-2020 and \citet{Burrows1997} models prefer a younger age for \epsindiba than for \epsindibb. To examine the statistical significance of this difference in model-derived ages between the two components we now consider only their measured masses and luminosities, excluding the rather uncertain stellar age.

\begin{figure*}
    \centering
    \includegraphics[width=0.32\textwidth]{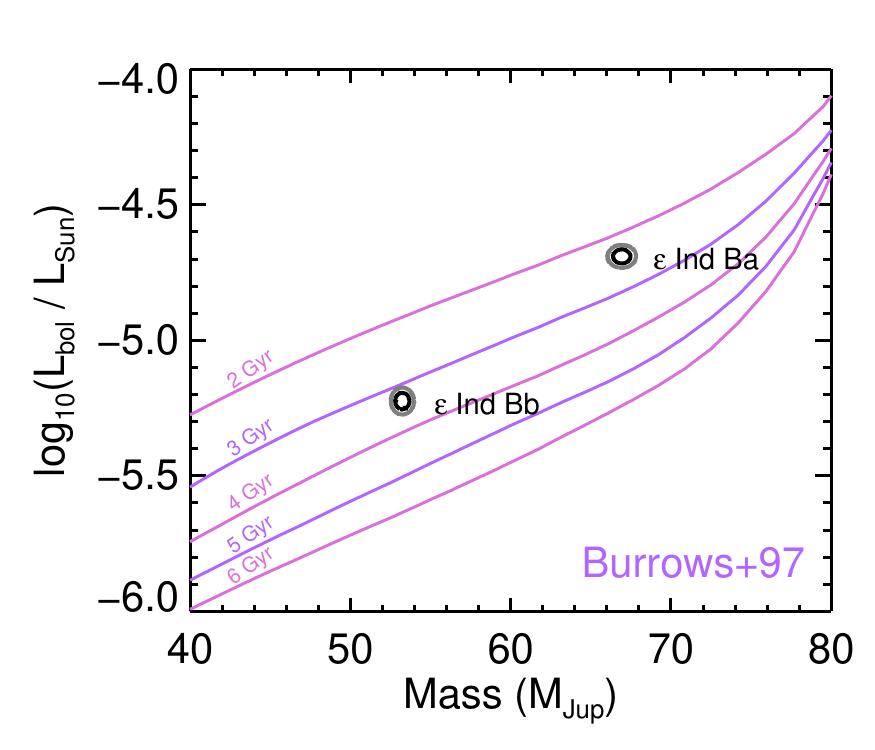}
    \includegraphics[width=0.32\textwidth]{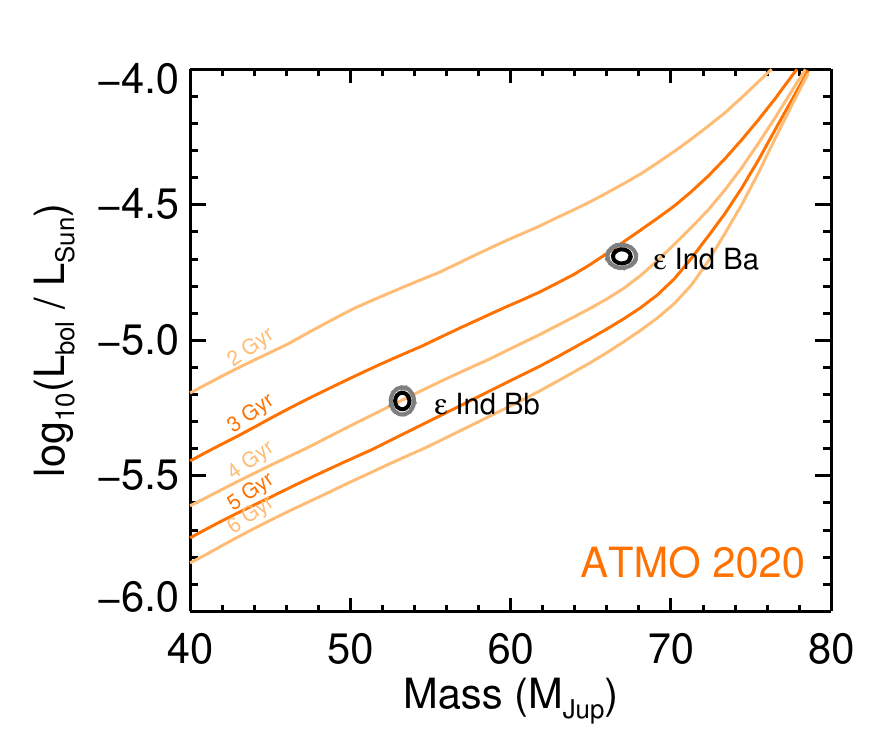}
    \includegraphics[width=0.32\textwidth]{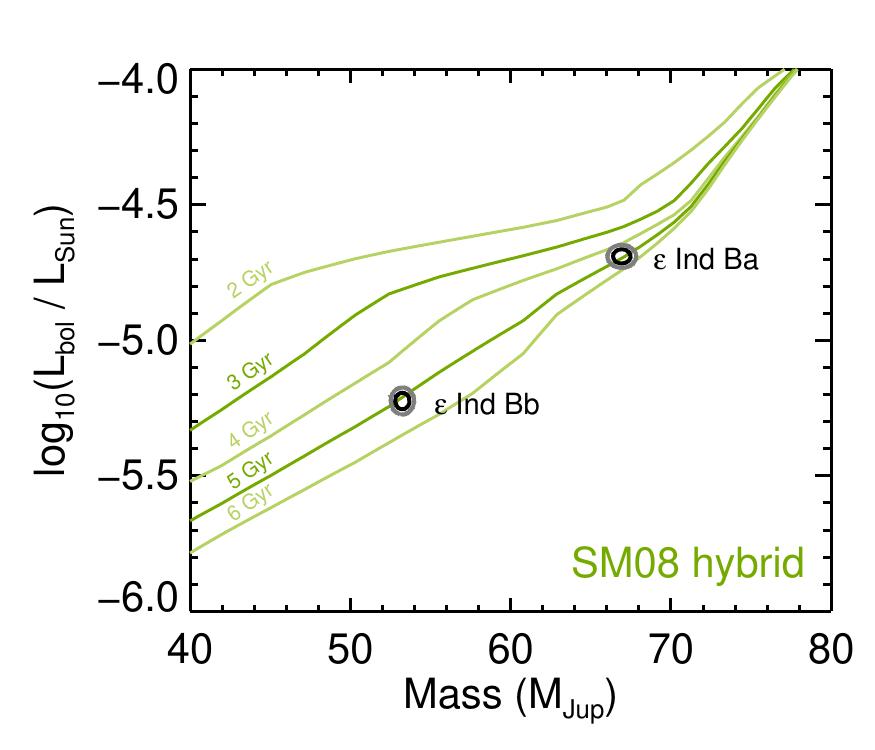}
    \caption{Isochrones from three different evolutionary models, ranging from 2\,Gyr to 6\,Gyr. Black and gray contours show the joint 1$\sigma$ and 2$\sigma$ confidence intervals of the masses and luminosities of \epsindiba and \Bb. Because these two brown dwarfs must be coeval they should lie along a single model isochrone. The only models that pass this test are the \citet{2008ApJ...689.1327S} hybrid models that predict a distinctly different mass--luminosity relation for brown dwarfs. These models have a much shallower dependence of luminosity on mass as objects cool through the L/T transition over $\Teff = 1400$\,K to 1200\,K, changing from cloudy to cloud-free atmosphere boundary conditions.}
    \label{fig:lbol-mass}
\end{figure*}

\begin{figure}
    \centering
    \includegraphics[width=0.45\textwidth]{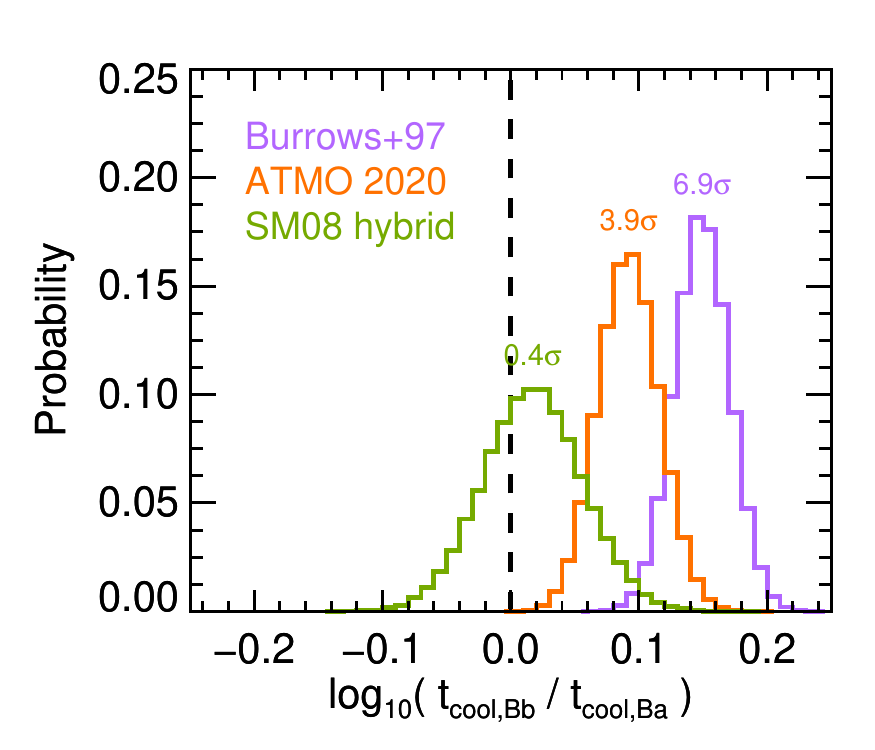}
    \caption{Probability distributions of the difference between the model-derived substellar cooling ages ($t_{\rm cool}$) of \epsindiba and \Bb. The dashed line shows the expectation that $t_{\rm cool, Bb} = t_{\rm cool, Ba}$. Only the \citet{2008ApJ...689.1327S} hybrid models predict consistent, coeval ages. This is the highest-precision coevality test of brown dwarf binaries to date, and it supports previous results from brown dwarf binaries with mass errors of $\approx$5\% \citep{2015ApJ...805...56D,Dupuy+Liu_2017}.}
    \label{fig:benchmark-coeval}
\end{figure}

\subsection{Isochrone test of $M$--$\Lbol$ relation for T~dwarfs}

Evolutionary models of brown dwarfs, from some of the earliest theoretical calculations up to modern work \citep[e.g.,][]{1993RvMP...65..301B,2020A&A...637A..38P}, typically predict a power-law relationship between mass and luminosity with a slope of $\Delta\log{L} / \Delta\log{M} = 2.5$--3.0. This general agreement between models with very different assumptions---and that vary greatly in other predictions such as the mass of the hydrogen-fusion boundary---can be seen in the slopes of isochrones for 40--60\,\Mjup\ brown dwarfs in Figure~\ref{fig:lbol-mass}.

One set of models, from \citet{2008ApJ...689.1327S}, that substantially alters the atmospheric boundary condition as objects cool from $\Teff = 1400$\,K to 1200\,K predicts a much shallower slope from the $M$--$L$ relation during that phase of evolution (Figure~\ref{fig:lbol-mass}). These so-called hybrid models provide the best match to the $M$--$L$ relation as measured in binaries composed of late-L dwarf primaries ($\Teff \approx 1400$\,K) and early-T dwarf secondaries ($\Teff \approx 1200$\,K), objects that straddle this evolutionary phase \citep{2015ApJ...805...56D,Dupuy+Liu_2017}. A fundamental prediction of these models is that during the L/T transition, objects of similar luminosity can have wider-ranging masses than in other models. The other chief prediction is that luminosity fades more slowly during the L/T transition, so that the brown dwarfs emerging from this phase are more luminous than in other models.

\epsindib is the only example of a binary with precise individual masses where one component is an L/T transition brown dwarf and the other is a cooler T~dwarf. This provides a unique test of the $M$--$L$ relation, where the cooler brown dwarf is well past the L/T transition and the other is in the middle of it. According to hybrid models, the brown dwarf within the L/T transition will be experiencing slower cooling, so it would be more luminous than in other models. On the other hand, with the immediate removal of cloud opacity in hybrid models below 1200\,K, a brown dwarf will cool even faster than predicted by other, non-hybrid models. These two effects predict that a system like \epsindib will, in fact, have an especially steep $M$--$L$ relation.

Our measured masses give a particularly steep slope for the $M$--$L$ relation of $\Delta\log{L} / \Delta\log{M} = 5.37\pm0.08$ between the L/T-transition primary \epsindiba and the cooler secondary \epsindibb. The only evolutionary models that predict such a steep slope are the hybrid models of \citet{2008ApJ...689.1327S}.

To quantitatively test models, we compared the model-derived cooling ages of \epsindiba and \Bb, given their measured masses and luminosities (Figure~\ref{fig:benchmark-coeval}). Models like ATMO-2020 that assume a single, cloud-free atmospheric boundary condition are 3.9$\sigma$ inconsistent with our measurements. At 6.9$\sigma$, models from \citet{Burrows1997} are even more inconsistent because the bunching up of isochrones around the end of the main sequence, which has a similar effect as the bunching up of isochrones due to slowed cooling in the hybrid models, occurs at higher masses than in ATMO-2020.

The \epsindib system therefore provides further validation of hybrid evolutionary models, where the atmosphere boundary condition is changed drastically over the narrow range of \Teff\ corresponding to late-L and early-T dwarfs. No longer just within the L/T transition, but affirming the consequences of slowed cooling during the L/T transition to cooler brown dwarfs ($\Teff < 1000$\,K).

\subsection{Testing Model Atmospheres: \Teff\ and \logg}

Brown dwarfs that have both directly measured masses and individually measured spectra have long been used in another type of benchmark test that tests for consistency between evolutionary models and the atmosphere models that they use as their surface boundary condition. Comparison of model atmospheres to observed spectra allows for determinations of \Teff, \logg, and metallicity. Evolutionary models predict brown dwarf radii as a function of mass, age, and metallicity. Combining these radii with empirically determined luminosities produce mostly independent estimates of $\Teff = (\Lbol/4\pi R^2 \sigma_{\rm SB})^{1/4}$, and with masses gives estimates of $\logg = \log(GM/R^2)$. (Evolutionary model radii have a small dependence on the model atmospheres and thus estimates of \Teff\ and \logg\ from their radii are not strictly, completely independent.) There are many examples of such benchmark tests, ranging from late-M dwarfs \citep[e.g.,][]{2001ApJ...554L..67K,2004ApJ...615..958Z,2010ApJ...721.1725D}, L~dwarfs \citep[e.g.,][]{2004A&A...423..341B,2009ApJ...692..729D,2010ApJ...711.1087K}, and T~dwarfs \citep[e.g.,][]{2008ApJ...689..436L,Dupuy+Liu_2017}.

From \citet{2010A&A...510A..99K}, \epsindiba and \Bb have perhaps the most extensive and detailed spectroscopic observations (0.6--5.1\,$\mu$m at up to $R\sim5000$) of any brown dwarfs with dynamical mass measurements. They found that BT-Settl atmosphere models \citep{2012RSPTA.370.2765A} with parameters of $\Teff = 1300$--1340\,K and $\logg = 5.50$\,dex best matched \epsindiba. For \epsindibb, they found $\Teff = 880$--940\,K and $\logg = 5.25$\,dex.

We computed evolutionary model-derived values for \Teff\ and \logg\ to compare to the model atmosphere results of \citet{2010A&A...510A..99K}. The most precise estimates result from using the mass and luminosity to derive a substellar cooling age and then interpolating \Teff\ and \logg\ from the same evolutionary model grid using the measured mass and the cooling age. The SM08 hybrid models gave $\Teff = 1312\pm13$\,K and $972\pm13$\,K for \epsindiba and \Bb, respectively, and $\logg = 5.365\pm0.006$\,dex and $5.288\pm0.003$\,dex. These evolutionary model-derived values agree remarkably well with the model atmosphere results, which were based on an atmosphere grid with discrete steps of 20\,K in \Teff\ and 0.25\,dex in \logg. 

ATMO-2020 models are only strictly appropriate for \epsindibb, and they give $\Teff = 992\pm13$\,K and $\logg = 5.311\pm0.003$\,dex. This effective temperature is $\approx$4$\sigma$ higher than the BT-Settl model atmosphere temperature. ATMO-2020 models are actually based on this family of model atmospheres (BT-Cond and BT-Settl should be effectively equivalent at this \Teff), so this suggests a genuine $\approx$50\,K discrepancy between atmosphere model-derived \Teff\ (too low) and evolutionary model-derived \Teff\ (too high). If so, this could be due a to a combination of systematics in atmosphere models (e.g., non-equilibrium chemistry, inaccurate opacities) and/or ATMO-2020 evolutionary model radii (10--20\% too high).

\section{Conclusions} \label{sec:conclusions}

In this paper we use $\sim$12 years of VLT data to infer dynamical masses of \MBa\,$\Mjup$ and \MBb\,$\Mjup$ for the brown dwarfs \epsindiba and \Bb, respectively. These masses put the the two objects firmly below the hydrogen burning limit. Our system mass agrees with that in \cite{Cardoso_2009}, who estimated a system mass of $121 \pm 1 \; M_{\rm Jup}$. With extra data from the completed relative and absolute astrometry monitoring campaign, we are able to derive precise individual masses and improve upon their previous analysis on several fronts. Using Gaia EDR3, we provide a much more precise calibration of both relative and absolute astrometry.  In addition, we have shown that our joint PSF fitting method accounts for the effect of overlapping halos reasonably well and adjusted our final errors for the relative astrometry according to our systematics analysis. Lastly, we have investigated and corrected for the systematics due to differential atmospheric refraction and residual atmospheric dispersion. As a result, we are able to obtain very tight constraints on the orbital parameters and final masses, and measure a parallax consistent with both the Hipparcos and Gaia values.

Our results disagree with \cite{Dieterich_2018}, who used the photocenter's orbit together with three NACO epochs to derive a mass of $75.0 \pm 0.82 \; M_{Jup}$ for \Ba, and a mass of $70.1 \pm 0.68 \; M_{Jup}$ for \Bb. These masses are at the boundaries of the hydrogen burning limit, challenging theories of substellar structure and evolution. We cannot conclusively say why \cite{Dieterich_2018} derive much higher masses.  However, we are able to reproduce their results, and find that rotating their measurements into an azimuth-only frame produces a mass closer to ours.  We speculate that highly asymmetric uncertainties in RA/Decl.~for a few of their measurements had a disproportionate effect on the results. 

We also provide a Fourier analysis of \epsindib's fluxes to investigate its potential variability. We find no definitive evidence of variability with a frequency less than 1 hr$^{-1}$.

Our newly precise masses and mass ratios enable new tests of substellar evolutionary models.  We find that \epsindiba and \Bb are generally consistent with cooling models at the activity age of $3.5^{+0.8}_{-1.0}$\,Gyr we derive for \epsindi~A.  However, the two brown dwarfs are consistent with coevality only under hybrid models like those of \cite{2008ApJ...689.1327S}, with a transition to cloud-free atmospheres near the L/T transition. 

Our masses for \epsindiba and \Bb, precise to $\approx$0.5\%, and our mass ratio, precise to $\approx$0.2\%, establish the \epsindib binary brown dwarf as a definitive benchmark for substellar evolutionary models.  As one of the nearest brown dwarf binaries, it is also exceptionally well-suited to detailed characterization with future telescopes and instruments including JWST.  \epsindib, with its two components straddling the L/T transition, now provides some of the most definitive evidence for cloud clearing and slowed cooling in these brown dwarfs. \\

\acknowledgements{T.D.B.~gratefully acknowledges support from National Aeronautics and Space Administration (NASA) under grant 80NSSC18K0439 and from the Alfred P.~Sloan Foundation.  MJM and CVC would like to thank their collaborators on the original ESO VLT NACO and FORS2 programmes which provided the great majority of the \epsindib astrometry data re-reduced and analysed in this paper: Laird Close, Ralf-Dieter Scholz, Rainer Lenzen, Wolfgang Brandner, Nicolas Lodieu, Hans Zinnecker, Rainer Köhler, and Quinn Konopacky. We would also like to recognise the tremendous efforts made by the many ESO service mode astronomers in carrying out these observations over many runs between 2004 and 2016, more than a full period of the binary orbit, and to the ESO TAC for continuing to support the programme throughout that time. Based on observations collected at the European Southern Observatory under ESO programs 072.C-0689(F), 073.C-0582(A), 074.C-0088(A), 075.C-0376(A), 076.C-0472(A), 077.C-0798(A), 078.C-0308(A), 079.C-0461(A), 380.C-0449(A), 381.C-0417(A), 382.C-0483(A), 383.C-0895(A), 384.C-0657(A), 385.C-0994(A), 386.C-0376(A), 087.C-0532(A), 088.C-0525(A), 089.C-0807(A), 091.C-0899(A), 381.C-0860(A), 072.C-0689(D), 075.C-0376(B), 076.C-0472(B), 077.C-0798(B), 078.C-0308(B), 079.C-0461(B), 380.C-0449(B), 381.C-0417(B), 382.C-0483(B), 383.C-0895(B), 384.C-0657(B), 385.C-0994(B), 386.C-0376(B), 087.C-0532(B), 088.C-0525(B), 089.C-0807(B), and 091.C-0899(B).}

\software{photutils (\citealt{Stetson_1987, photutils110}), astropy (\citealt{Astropycollab2013aaAstropy}), orvara (\citealt{orvara_2021}), Scipy (\citealt{2020SciPy-NMeth}), matplotlib (\citealt{Hunter:2007}), numpy (\citealt{harris2020array})}\\

\end{document}